\documentclass[12pt,preprint]{aastex}

\usepackage{color}
\usepackage{psfig}

\newcommand{\water}{\mbox{${\rm H}_{2}{\rm O}$}}
\newcommand{\app}{\ensuremath{\sim} }

\newcommand{\wwater}{H$_2$O}

\shorttitle{Cooling of Dense Gas by H$_2$O Line Emission}
\shortauthors{Morris et al.}

\begin{document}

\title{Cooling of Dense Gas by H$_2$O Line Emission and an Assessment of its Effects in Chondrule-Forming Shocks}

\author{M.~A.~Morris and S.~J.~Desch}
\affil{School of Earth and Space Exploration, Arizona State University,
        P.~O.~Box 871404, Tempe, AZ 85287-1404}

\and

\author{F.~J.~Ciesla\altaffilmark{1}}
\affil{Department of Terrestrial Magnetism, Carnegie Institution of    
         Washington, 5241 Broad Branch Road NW, Washington DC 20015-1305}
\email{melissa.a.morris@asu.edu}

\altaffiltext{1}{Now at Department of the Geophysical Sciences, 
   University of Chicago, 5734 Ellis Ave., Chicago IL 60637}


\begin{abstract}

We consider gas at densities appropriate to protoplanetary disks and calculate its ability to cool due to line radiation emitted by $\water$ molecules within the gas.  Our work follows that of Neufeld \& Kaufman (1993; \apj, 418, 263), expanding on their work in several key aspects, including use of a much expanded line database, an improved escape probability formulism, and the inclusion of dust grains, which can absorb line photons.  Although the escape probabilities formally depend on a complicated combination of optical depth in the lines and in the dust grains, we show that the cooling rate including dust is well approximated by the dust-free cooling rate multiplied by a simple function of the dust optical depth.  We apply the resultant cooling rate of a dust-gas mixture to the case of a solar nebula shock pertinent to the formation of chondrules, millimeter-sized melt droplets found in meteorites.  Our aim is to assess whether line cooling can be neglected in chondrule-forming shocks or if it must be included.  We find that for typical parameters, $\water$ line cooling shuts off a few minutes past the shock front; line photons that might otherwise escape the shocked region and cool the gas will be absorbed by dust grains.  During the first minute or so past the shock, however, line photons will cool the gas at rates $\sim 10^4 \ {\rm K} \, {\rm hr}^{-1}$, dropping the temperature of the gas (and most likely the chondrules within the gas) by several hundred K.  Inclusion of $\water$ line cooling therefore must be included in models of chondrule formation by nebular shocks.  
\end{abstract}


\section{Introduction}

\subsection{Radiative Transfer and Line Cooling}

In many astrophysical settings, emission of radiation via
rotational and vibrational transitions of molecules plays
an important role in cooling warm gas.
Because the emission of such radiation is in sharp spectral
lines, this mechanism is referred to as line cooling.
Line radiation from the water molecule ${\rm H}_{2}{\rm O}$,
with its permanent electric dipole and its high cosmochemical
abundance, is significant in a variety of settings ranging
from molecular clouds to protostellar envelopes (e.g.,
Cernicharo \& Crovisier 2005).
More recently, line cooling from ${\rm H}_{2}{\rm O}$ molecules
has been recognized to play a pivotal role in the energetics 
following the passage of a shock wave through the dense gas in
the solar nebula protoplanetary disk (see Desch et al.\ 2005).

The rate of emission of line radiation from a warm gas containing
water molecules is difficult to calculate, because there are 
so many accessible rotational and vibrational energy levels,
and therefore a great many transitions.
In dense molecular gas ($n_{\rm H2} > 10^{10} \, {\rm cm}^{-3}$),
these energy levels are populated according to Boltzmann statistics
(i.e., in local thermodynamic equilibrium, or LTE), but at lower 
densities the populations must be calculated by balancing 
transition rates.
Finally, the cooling of gas by line radiation relies on the ability
of the photons to escape the system without being reabsorbed; to
the extent that other, nearby molecules can absorb the emitted
photons, the gas does not cool.
Photons generated during a molecular transition are emitted
and also reabsorbed over a small range of frequencies centered 
on the line frequency, with an efficiency that depends on the
frequency shift from the line center.
Rather than integrating the equations of radiative transfer 
over all frequencies within the line, all standard treatments of 
the problem instead assume integration over this ``line profile" 
and use a frequency-integrated escape probability of photons.
This escape probability is a function of the column density of 
the molecule within the system.
This is the approach taken by Neufeld \& Kaufman (1993; hereafter NK93),
in particular.

The calculation of NK93 of the rate of line cooling from ${\rm H}_{2}{\rm O}$
molecules has remained the state of the art for 15 years; but, as we
discuss below, several aspects of the NK93 calculation are now out of date.
Their calculation uses a limited database of transitions,
an oversimplified escape probability formulism, and ignores absorption
by dust grains.
It is one of the goals of the work here to update the ${\rm H}_{2}{\rm O}$
line cooling rates to improve on the calculations of NK93.
A second goal of this paper is assess whether or not line cooling plays a
significant role in cooling the gas following the passage of a shock
through the dense gas of the solar nebula.
As discussed by Desch et al.\ (2005), previous modeling has not
determined whether or not line cooling can be neglected.
This has significant implications for the issue of chondrule formation.

\subsection{Chondrules}

The parent bodies of the most primitive meteorites, the chondrites, formed at about 
2-3 astronomical units (AU) from the Sun, 4.57 billion years ago 
(Wadhwa \& Russell 2000).
Chondrites are the 
most primitive meteorites in our collection, in that they have suffered very
little alteration since their formation, and therefore contain information about the 
conditions that existed in the early solar nebula.  
Chondrites are remarkable for containing calcium-rich, aluminum-rich inclusions (CAIs),
the oldest solids in the Solar System, whose formation has been dated to 
$4567.2 \pm 0.6$ Mya (Amelin et al.\ 2002).
Also found in abundance within chondrites are sub-millimeter- to millimeter-sized, (mostly ferromagnesian) 
igneous spheres, called chondrules, from which the chondrites derive their name. 
Chondrules formed, at most, \app~2-3 million years after CAIs (Amelin et al. 2002; 
Kita et al. 2005), as melt droplets that were heated to high temperatures while they were 
independent, free-floating objects in the early solar nebula.  
After they were heated, cooled and recrystallized, chondrules were incorporated into the 
parent bodies from which chondrites originate.  
Chondrules are capable of providing incredibly detailed information about conditions
in the Solar System protoplanetary disk, if the process that led to their heating, melting
and recrystallization could be understood.
Chondrules make up to 80\% of the volume of ordinary chondrites (Grossman 1988), and it is 
estimated that \app 10$^{24}$ g of chondrules exist in the asteroid belt today (Levy 1988).  
Such a prevalence of chondrules suggests that chondrule-forming events were widespread in 
the solar nebula.  
A process that can melt $10^{24} \, {\rm g}$ of rock is surely a dominant process in the 
solar nebula disk.

The chondrule formation process, despite its obvious importance, has been a mystery in
the field of meteoritics for two centuries (Sorby 1877).
Proposed mechanisms include interaction with the early active Sun, through jets (Liffman 
\& Brown 1995; Liffman \& Brown 1996) or solar flares (Shu et al. 1996, 1997, 2001),
melting by lightning (Pilipp et al.\ 1998; Desch \& Cuzzi 2000), and melting by planetesimal
impacts (Urey \& Craig 1953; Urey 1967; Sanders 1996; Lugmair \& Shukolyukov 2001).
The dominant theory, though, is that chondrules were melted in shock waves in the 
protoplanetary disk (Hewins 1997; Jones et al. 2000; Connolly \& Desch 2004; Connolly et al. 2006).
A shock wave, or shock front, is a sharp discontinuity between supersonic and subsonic gas, 
over an area only a few molecular mean free paths thick, typically only meters in the solar 
nebula.
The gas is slowed, compressed, and heated by the time it reaches the other side of the shock 
front.  
Solids moving with the gas are heated not only by thermal exchange upon entering the shocked 
region, but also by friction as they are slowed to the post-shock gas speed, and by absorbing 
the infrared radiation emitted by other solids.
The shock model of chondrule formation appears able to resolve the chondrule formation
mystery, because it makes several detailed predictions about chondrule formation that 
are largely borne out by observation and experimentation, especially regarding chondrule
thermal histories. 

\subsection{Chondrule Formation by Nebular Shocks}

Heating and cooling rates of chondrules have been determined experimentally.
According to furnace experiments, in which melt droplets with chondrule compositions
are allowed to cool and crystallize, reproduction of chondrule textures requires 
particular ranges of cooling rates between the liquidus temperature ($\approx 1800$ K)
and solidus temperature ($\approx 1400$ K; Hewins \& Connolly 1996).
Porphyritic  chondrules cooled at about $10 -10^3 \, {\rm K} \, {\rm hr}^{-1}$, and 
barred-olivine chondrules cooled at about $10^3 \, {\rm K} \, {\rm hr}^{-1}$ (Hewins et al.\
2005; see also Desch \& Connolly 2002).
Additionally, chondrules retain volatile elements such as S, indicating that they did not 
remain above the liquidus for more than minutes, and cooled quite rapidly 
($\gtrsim 10^4 \, {\rm K} \, {\rm hr}^{-1}$ while above the liquidus (Yu \& Hewins 1998)). 
Finally, there is no indication of the isotopic fractionation that would arise from the
free evaporation of alkalis such as Na, which 
constrains the time spent at high temperature before melting (Tachibana et al. 2004). 
Modeling of isotopic fractionation has shown that chondrules must heat up from 1300 to 1600 K 
on the order of minutes or less (Tachibana \& Huss 2005).

Passage through nebular shocks satisfies nearly all the experimental constraints on chondrule 
formation in broad brush (Iida et al. 2001, hereafter INSN; Desch \& Connolly 2002, hereafter 
DC02; Ciesla \& Hood 2002, hereafter CH02; Desch et al. 2005).  
Prior to passage through the shock front, chondrules are moderately heated by absorbing 
radiation emitted by chondrules which have already passed through the shock front. 
Upon passage through the shock front, the gas is immediately slowed, and compressed and 
heated, but the chondrules continue at supersonic speeds through the gas.
They achieve their peak heating during this stage, due to absorption of radiation, 
thermal exchange with the gas, and supersonic frictional drag heating. 
This stage lasts until the chondrules slow to the gas velocity, which takes an
aerodynamic stopping time, $t_{\rm stop} = \rho_{\rm s} a_s / \rho_{\rm g} C \sim 1$ minute
(where $\rho_{\rm s}$ and $a_s$ are the particle density and radius, and 
$\rho_{\rm g}$ and $C$ are the post-shock gas density and sound speed).
At this time, the gas and chondrules are dynamically coupled and chondrules are heated
only by absorption of radiation and thermal exchange with the gas.
Soon thereafter the solids and gas become thermally coupled as well, and both
components achieve similar temperatures.  
Although the chondrules pass through the first 100 km of the post-shock region more rapidly 
than the gas, once they do, they will thermally equilibrate to the gas temperature.  
Once this happens, the gas and chondrules cool together as fast as they can either radiate 
away energy or leave the source of infrared radiation that heats them.
DC02 showed that the cooling rate of gas and chondrules should be 
\begin{equation}
CR=50 \; \left( \frac{\rho_{\rm g}}{10^{-9} {\rm g} \, {\rm cm}^{-3} } \right) 
         \left( \frac{V_{\rm s}}{7 \, {\rm km} \, {\rm s}^{-1}}  \right) 
         \left( \delta + \frac{C}{50} \right) \, {\rm K} \, {\rm hr}^{-1},
\label{eq:cr}
\end{equation}
where $\rho_{\rm g}$ is the pre-shock gas density, $V_{\rm s}$ the shock speed,
$C$ is the ``concentration" of chondrules
(the chondrules-to-gas mass ratio normalized to $3.75 \times 10^{-3}$,
where a chondrule radius $300 \, \mu\rm m$ has been assumed),
and $\delta$ the concentration of submicron dust
(the dust-to-gas mass ratio normalized to $1.25 \times 10^{-3}$).
More precisely, $\delta$ should be interpreted as the Rosseland mean opacity of the dusty gas, 
normalized to $1.14 \, {\rm cm}^{2} \, {\rm g}^{-1}$.
For typical parameters, then, cooling rates $\sim 10^2 \, {\rm K} \, {\rm hr}^{-1}$
are obtained, consistent with the cooling rates of chondrules. 

\subsection{Estimates of Line Cooling in Shocks}

The effect of line cooling---the cooling of gas by emission of so-called line
photons by trace molecules in the gas such as CO and ${\rm H}_{2}{\rm O}$---
in solar nebula shocks has been considered by INSN and by Miura \& Nakamoto (2006).
In both cases the cooling rates of NK93 
were used.
INSN assumed a gas optically thin to the line radiation and found chondrule
cooling rates $\sim 10^{4} \, {\rm K} \, {\rm hr}^{-1}$ in many cases.
Miura \& Nakamoto (2006) allowed the gas to become optically thick to this
line radiation, using the formulation of NK93 described below. 
They found that for plausible parameters the cooling rates of chondrules and 
gas cluster around $5000 \, {\rm K} \, {\rm hr}^{-1}$, with line cooling playing an important
role in the cooling of the gas. 
Chondrules and gas are thermally coupled only a few minutes after passing through
the shock front (INSN; DC02; CH02), so these chondrule cooling rates are 
controlled by the rate at which gas cools by emission of line radiation. 
To understand the essence of their results without reproducing all of their 
calculations in detail, we now estimate the cooling rate of chondrules and gas following 
the shock, using the tabulations of NK93.

We first examine how the cooling rates predicted by NK93 and used by INSN
and Miura \& Nakamoto (2006) are calculated. 
For a wide range of densities and temperatures, NK93 calculated the steady-state 
level populations of $\water$ and CO molecules (denoted `$M$'), and the rates of 
emission of photons.
The escape of these photons from a system was considered under the large-velocity-gradient 
(Sobolev) approximation as a function of column density $n(M) d$, via use of the parameter
\begin{equation}
\tilde{N} \equiv \frac{n(M)}{d v_{z} / d z}, 
\end{equation}
These results were adapted to a large range of geometries of interest, including those
not consistent with the Sobolev approximation, in particular the one of relevance to nebular 
shocks, a static, plane-parallel slab of thickness $d$.
According to NK93, the cooling rate at the center of the slab (at depth $d/2$) will depend on the similar 
parameter
\begin{equation}
\tilde{N} = \frac{n(M)\;d}{\Delta v},
\end{equation}
where $\Delta v$ is the thermal velocity of the molecules $M$.
An important assumption made by NK93 is that if the line photons can escape
to the edge of the slab, then the gas is cooled; if the photons do not escape
to the edge of the slab, they are essentially re-absorbed on the spot and there
is no cooling. 
The cooling rate per volume is then $n(m) {\cal L}_{\rm LTE} (\tilde{N})$,
where the subscript `LTE' is the notation used by NK93 to denote cooling rates derived
under the assumption that the population levels follow a Boltzmann distribution, 
which is the regime of interest for solar nebula shocks.
NK93 have tabulated ${\cal L}_{\rm LTE}(\tilde{N})$ for both vibrational and 
rotational lines of $\water$ and CO, for every decade in $\tilde{N}$.  

The plane-parallel slab is equivalent to the nebula shock geometry because of
the large velocity jump at the shock front.  
In a typical chondrule-forming shock, the gas velocity changes nearly 
instantaneously from $\approx 7 \, {\rm km} \, {\rm s}^{-1}$ to 
$\approx 1.6 \, {\rm km} \, {\rm s}^{-1}$, a jump of over 
$5 \, {\rm km} \, {\rm s}^{-1}$. 
This velocity difference is much larger than the thermal velocities of the
$\water$ molecules [$\Delta v = (2 k T / m_{\rm H2O})^{1/2}$
$= 1.36 \, {\rm km} \, {\rm s}^{-1}$ at 2000 K] that control the width of the
line profile.
Essentially, once a line photon emitted from the post-shock gas escapes to the 
shock front, it will continue to travel relatively unimpeded by gas. 
Only one change to the NK93 formulation must be made before it is adopted to 
the nebular shock geometry.
For gas a distance $z$ past the shock front, the equivalent thickness of the 
slab is $d = 2 z$, and the cooling rate is only half the cooling rate calculated
by NK93, who assumed radiation could escape either side of the slab.

The cooling rate of the gas yields 
\begin{equation}
\frac{\partial e}{\partial t} = 
\frac{\partial}{\partial t} \left( \frac{p}{\gamma -1} \right) = 
2.8 \, n_{\rm H2} \, k \, \frac{\partial T}{\partial t} =
- \frac{1}{2} \; n(\water)\;\mathcal{L_{\rm LTE}}(\tilde{N}) ,
\end{equation}
where we have ignored compression of the gas after passing through the
shock front in this toy model, and we have assumed an abundance of He of 10\% by number.
Converting the time derivative to a spatial derivative by assuming a constant
gas velocity $V_{\rm g}$ past the shock front, we find 
\begin{equation}
\frac{\partial T}{\partial \tilde{N}} = -\frac{ \Delta v }{ 5.6 n_{\rm H2} V_{\rm g} k } \,
 \frac{ {\cal L_{\rm LTE}}(\tilde{N}) }{ 2 },
\label{eq:nkcool}
\end{equation}
where we have assumed a constant $\Delta v$ throughout the post-shock region.
In actuality, $\Delta v$ would decrease past the shock front, but that would
merely reduce the total cooling, so our assumption overestimates the degree of
cooling. 
This formula provides the gas temperature as a function of $\tilde{N}$, the 
effective column density past the shock front.

Fedkin \& Grossman (2006) showed that the $n_{\rm H2O} / n_{\rm H2}$ ratio in the solar nebula probably varied with temperature, but found that $n_{\rm H2O} / n_{\rm H2} = 5 \times 10^{-4}$ was a good average for the temperatures of interest. 
We have adopted a canonical ratio $n_{\rm H2O} / n_{\rm H2} = 8 \times 10^{-4}$.
This is only slightly higher than some commonly assumed canonical ratios but is consistent with the abundance  of Lodders (2003), $n_{\rm H2O} / n_{\rm H2} = 8.88 \times 10^{-4}$.  
We note that the exact value of $n_{\rm H2O} / n_{\rm H2}$
in the chondrule formation environment is likely variable with time (Ciesla \& Cuzzi 2006),
and consider variations in this ratio later in the paper.
Some dissociation of $\water$ is likely to occur at the peak post-shock temperatures we 
consider, but we are interested in the maximum cooling possible by line photons, so we neglect 
dissociation in our calculations. 
 
Interpolating between the tabulated values of $\mathcal{L}_{\rm LTE}(\tilde{N})$ provided by 
NK93, we have integrated equation~\ref{eq:nkcool} to find the total drop in temperature
by the time the gas is at an equivalent of 
$\tilde{N} = 10^{21} \, {\rm cm}^{-2} \, {\rm km}^{-1} \, {\rm s}$
past the shock front.
For typical parameters (pre-shock density $n_{\rm H2} = 2 \times 10^{14} \, {\rm cm}^{-3}$
and velocity $V_{\rm g} = 7 \, {\rm km} \, {\rm s}^{-1}$; mixing ratio 
$n_{\rm H2O} / n_{\rm H2} = 8 \times 10^{-4}$; post-shock density 
$n_{\rm H2} = 1.2 \times 10^{15} \, {\rm cm}^{-3}$ and velocity 
$V_{\rm g} = 1.2 \, {\rm km} \, {\rm s}^{-1}$; and post-shock thermal
velocity $\Delta v = 1.6 \, {\rm km} \, {\rm s}^{-1}$), 
this equates to $d = 1.3 \times 10^{4} \, {\rm km}$, a distance
$z = 6.6 \times 10^{3} \, {\rm km}$, and a time 1.6 hours after passing through 
the shock front.
(Again, this toy model neglects the compression and slowing of the gas.
It also assumes wrongly that $\Delta v$ is constant, whereas it should decrease
slightly as the gas cools.  It is nonetheless illustrative.)
At these distances past the shock front, the total cooling due to rotational lines of
$\water$ and CO is completely negligible, $< 4 \, {\rm K}$.  
This illustrates that the rotational lines become optically thick so quickly that
they are unable to escape the post-shock region and cool the gas.
On the other hand, the total cooling from CO vibrational line photons is roughly
72 K.  
This leads to an average cooling rate of $\sim 40 \, {\rm K} \, {\rm hr}^{-1}$,
which is important, but not significantly greater than the cooling rates of chondrules 
($\sim 10^{2} \, {\rm K} \, {\rm hr}^{-1}$).
The assumption of INSN that the lines of $\water$ and CO are 
optically thin is therefore invalidated, because these lines are indeed optically 
thick.  The final line cooling mode (not considered by INSN, but considered by Miura \&
Nakamoto 2006) is vibrational line cooling from $\water$.   
We find a total cooling due to these line photons on the order of 800 K.
This is a significant cooling of the gas: the average cooling over 1.6
hours is $\sim 500 \, {\rm K} \, {\rm hr}^{-1}$, with even higher cooling rates 
obtained early on.
The cooling of the gas by vibrational line photons of $\water$, despite being 
somewhat optically thick, is the source of the high chondrule cooling rates
predicted by Miura \& Nakamoto (2006).  We confirm the statement made by Miura \& Nakamoto (2006) that ``...line emission is important for the gas cooling.''  
We hereafter focus on cooling due to $\water$ alone, as the dominant coolant is expected to be water. 

Clearly, any model of chondrule cooling rates in solar nebula shocks must
account for the emission of line photons from $\water$. 
However, incorporating the cooling rates of NK93 directly into a shock code,
as Miura \& Nakamoto (2006) did, is not ideal.
First, line cooling due to rotational and vibrational transitions of $\water$ was 
calculated by NK93 using roughly 50,000 transitions from the HITRAN database 
(Rothman et al. 1987).
While this was the best available data at the time, much more extensive databases
now exist.
Second, NK93 calculated escape probabilities under the large-velocity-gradient (Sobolev) 
approximation, and asserted that the results will apply to the case of a static, plane-parallel 
slab if $\tilde{N}$ is defined as described above. 
Third, NK93 assumed a one-sided escape probability for line photons equal to 
$0.5 / (1 + 3\tau)$, where $\tau$ is the Sobolev optical depth; more exact escape 
probabilities exist, as described below.
Fourth, it is not necessarily the case that if photons do not escape to the 
edge of the slab that the gas does not cool locally; photons re-absorbed halfway
to the edge still cool the gas at the slab center.
Finally, NK93 did not consider the absorption of line photons by intervening dust.
Inclusion of dust is absolutely mandatory, as the following argument makes clear: 
For a solar-composition gas with 0.5\% of the mass in the form of $a = 0.5 \, \mu{\rm m}$ 
radius grains (similar to the size of matrix dust grains in meteorites), the opacity to 
short-wavelength radiation is as high as $\kappa = (\rho_{\rm dust} / \rho_{\rm gas}) 
\, (3 / 4 \rho_{\rm s} a) = 30 \, {\rm cm}^{2} \, {\rm g}^{-1}$, where the internal density
of the dust grains is $\rho_{\rm s} \approx 2.5 \, {\rm g} \, {\rm cm}^{-3}$ . 
Over the distance for which the optical depth in dust is unity, the parameter
$\tilde{N}$ is therefore less than $10^{19} \, {\rm cm}^{-2} \, {\rm km}^{-1} \, {\rm s}$,
assuming a ratio $n_{\rm H2O} / n_{\rm H2} = 8 \times 10^{-4}$.
This corresponds to distances $\sim 10^{3} \, {\rm km}$, or times of about 10 minutes,
after passage through the shock front. 
As this is smaller than the $\tilde{N}$ for which the most significant cooling takes place, 
dust grains are capable of absorbing photons that would otherwise cool the gas.
A thorough treatment of line cooling should improve on the calculations of
NK93 in the areas identified above.

\subsection{Outline}

In this paper we calculate the cooling rate of gas due to emission of line photons
by $\water$ molecules, improving upon the work of NK93 with the use of an expanded database, 
improved escape probabilities, and the inclusion of dust. 
First we evaluate escape probability approximations, both in the presence and absence of 
dust.
We then calculate the cooling rates from $\water$ molecules using the 1.2-million
line SCAN-$\water$ database (J\"{o}rgensen et al.\ 2001).
Finally, we incorporate these cooling rates into a toy model of chondrule thermal 
evolution.
We conclude that for canonical parameters, line cooling is significant in the 
first few minutes past the shock front, but that afterward dust grains absorb the 
line photons that would otherwise cool the gas. 
We discuss the effect of the important dust-to-water ratio on these conclusions.

\section{Calculation of $\water$ Cooling Rates} 

We have used the SCAN-$\water$ database of J{\o}rgensen et al. (2001) of 1.2 million lines 
(rotational plus vibrational) of the $\water$ molecule to calculate the cooling rates due to 
emission of line photons from $\water$.  
We continue to assume, as NK93 did, that if photons are not able to escape the region
entirely, they are re-absorbed on the spot. 
This assumption will be relaxed in future work, but for now it allows a crude estimate
of the importance of line radiation. 
In other respects, we improve on the NK93 calculation by more accurately 
calculating escape probabilities, and by including the possibility of absorption by dust 
grains.

\subsection{Escape Probabilities Without Dust}

The probability that a line photon will escape a semi-infinite volume, otherwise known as the one-sided 
escape probability, $P_{\rm esc}$, is given by   
\begin{equation}
P_{\rm esc} =  \frac{1}{2}\;\int^{\infty}_{-\infty}\Phi(x)\;E_{2}(\tau \Phi(x) )\;dx,
\label{eq:pesc} 
\end{equation}
\begin{equation}
x  \equiv \frac{\lambda-\lambda_0}{\Delta\lambda_D},
\label{eq1.1}  
\end{equation}
\begin{equation}
{\Delta\lambda_D} = \frac{c}{\lambda_0} \; \sqrt{\frac{2kT}{m_{\rm H2O}}}
\label{eq1.2} 
\end{equation}
(Avrett \& Hummer 1965; Hummer \& Rybicki 1971; Rybicki \& Lightman 1979; Bowers \& Deeming 1984).  
Here $\Phi$(x) is the line profile, $x$ is the frequency measured from line center in Doppler widths 
($\Delta \lambda_D$), and $E_{2}$ is the second exponential integral.
In equation~\ref{eq:pesc}, $\tau$ refers to the optical depth integrated over the line, which is $\sqrt{\pi}$ times the optical 
depth at line center. 
Considering the combined effect of both Doppler and Lorentz broadening, the line profile is given by the normalized 
Voigt profile 
\begin{equation}
\Phi(x)=\frac{a}{\pi^{3/2}}\;\int^{\infty}_{-\infty}\frac{e^{-y^{2}}}{(x-y)^{2}+a^{2}}\;dy,
\label{eq2} 
\end{equation}
(Avrett \& Hummer 1965; Rybicki \& Lightman 1979; Bowers \& Deeming 1984), where $a$ is the ratio of Lorentz to Doppler width 
(Avrett \& Hummer 1965).  
In astrophysical situations $a \ll 1$, perhaps as large as 0.1 (Mihalas 1978; Bowers \& Deeming 1984).
In this study, the Voigt profile was calculated using the algorithm of Zaghloul (2007), where the 
Voigt function is written as a single proper integral with a damped sine integrand.  
A straightforward numerical integration was performed to calculate the ``exact" one-sided escape 
probability $P_{\rm esc}$ in the absence of dust.

Besides the form adopted by NK93, several other approximations to the one-sided escape probability 
exist in the literature.     
Hollenbach \& McKee (1979) give the following approximation to the one-sided escape probability:
\begin{equation}
P_{\rm esc}=\frac{1}{2} \; \frac{1}{1 + \tau \left(2 \;\mbox{ln} \left( 2.13 + \tau^2/\pi \right) \right) ^ {1/2}}.
\label{eq9} 
\end{equation}
The approximation given by Collin-Souffrin et al. (1981) is
\begin{equation}
P_{\rm esc}=\frac{1}{2}\;\frac{1}{1+2\tau \left(\mbox{ln} \left( \tau / \sqrt{\pi}+1 \right) \right) ^{1/2}}.
\label{eq11} 
\end{equation}
Mathews (1992) estimates 
\begin{equation}
P_{\rm esc}=\frac{1}{2} \; \frac{1}{1+2 \tau \left( \mbox{ln} \left( \tau \sqrt{\pi} + 1 \right) \right) ^{1/2}},
\end{equation}
and Dumont et al. (2003) give
\begin{eqnarray}
P_{\rm esc} = \frac{1-e^{-2\tau}}{4 \tau}, \;\;\;\;\;\;\;\;\;\;\;\;\;\;\;\;\; & \tau < 1, \\
P_{\rm esc} = \; \frac{1}{2 \sqrt{\pi} \, \tau \left( 1.2 + {\rm ln} \frac{ \tau^{1/2} }{ 1 + 10^{-5} \tau } \right) }, & \tau > 1.
\end{eqnarray}
These approximations are all in the Doppler limit, in which $a = 0$ and the line profile is given by $\pi^{-1/2} \, e^{-x^2}$.  
We have evaluated the approximations to the escape probabilities of Hollenbach \& McKee (1979), Collin-Souffrin et al. (1981), 
Mathews (1992), NK93, and Dumont et al. (2003), and compared the results to our ``exact" escape probabilities (Figure 1).  
The maximum error between these approximations and the exact escape probability are as follows: 
11.42\% for Hollenbach \& McKee (1979); 
15.70\% for Collin-Souffrin et al. (1981);
31.34\% for Mathews (1992);
and 37.22\% for Dumont et al. (2003).  
The maximum error between the approximation used by NK93 and the exact escape probability is 46.44\%, although it is not quite 
appropriate to make this comparison, as NK93 used the Sobolov optical depth in their calculations. 
The approximation of Hollenbach \& McKee (1979) is superior to the others and provides a satisfactory fit; we will use this
approximation in what follows. 
\begin{figure}
\epsscale{0.8}   
\plotone{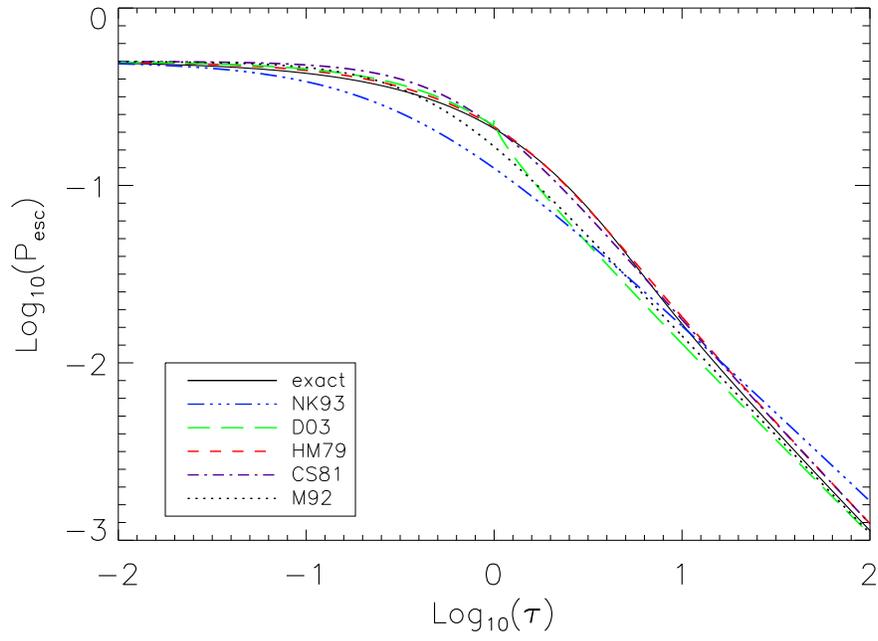}
\caption{Escape probability approximations used by Hollenbach \& McKee (1979), Collin-Souffrin et al. (1981), Mathews (1992), NK93, and 
Dumont et al. (2003), compared to the ``exact" escape probability calculated here (see text).}
\end{figure}

Using the Hollenbach \& McKee (1979) approximation to the escape probability, a parameter study was conducted to determine the effect 
of varying the value of $a$ in the calculation of the Voigt function.  
It was found that varying the value of $a$ between 0.0 and 0.1 increased the escape probability significantly for $\tau \gg 1$ (Figure 2), 
consistent with figures 4 and 5 of Hummer \& Rybicki (1982).  
It is not immediately clear how this will affect the overall cooling rate, but we anticipate a small effect as the lines at which the escape 
probability increases with increasing $a$ are optically thick, leading to lower cooling rates.  
We investigate this effect of the Voigt function on the cooling rates in Section 3.
\begin{figure}
\epsscale{0.8}         
\plotone{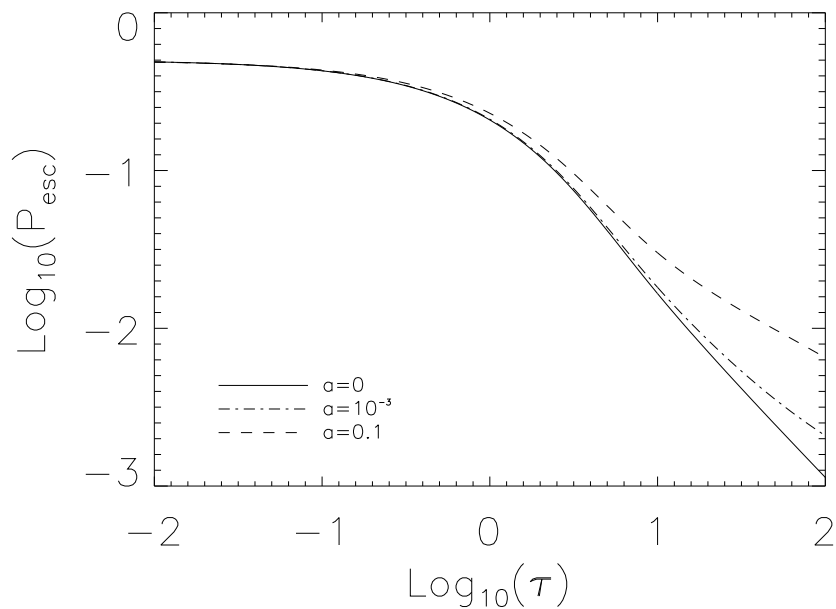}
\caption[esc_a.eps]{Exact escape probability for three different values of $a$ (the ratio of Lorentz to Doppler width).}
\end{figure}

\subsection{Escape Probabilities With Dust}

\begin{figure}
\epsscale{0.8}        
\plotone{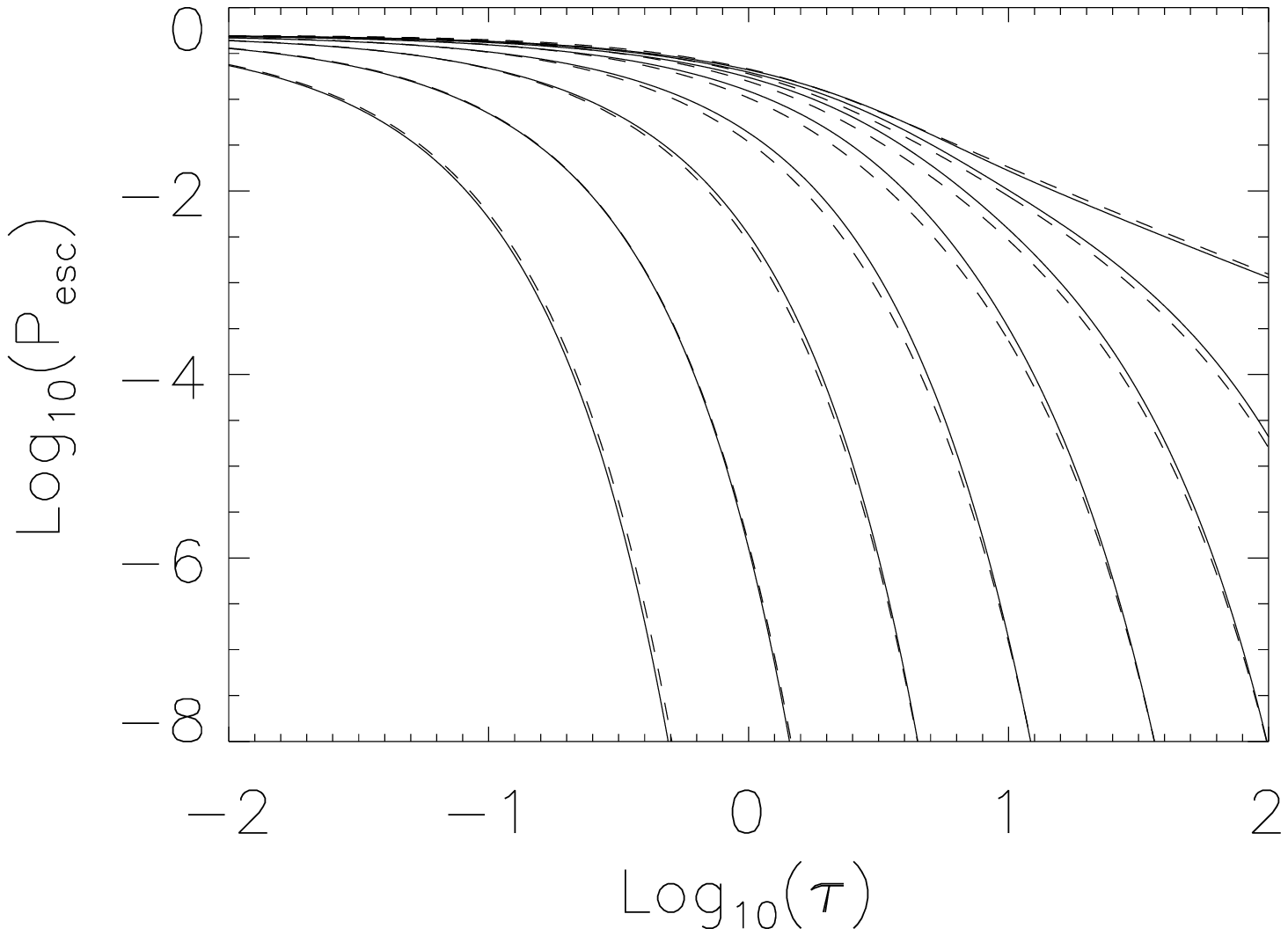}
\caption{Approximations to the escape probability with the inclusion of dust (dashed lines), compared to exact 
escape probabilities, including dust (solid lines).  
From left to right, the values used for $\tau_{\rm d} / \tau_0$ are 30, 10, 3, 1, 0.3, 0.1, 0.03, and 0.}
\label{fig:sprinkler} 
\end{figure}

Dust can also absorb photons, preventing their escape.  
Hence, the proper treatment of the escape probability must account for the absorption by dust, where 
the optical depth to line photons is now given by $\tau_d + \tau \Phi(x)$, and the exact escape probability is  
\begin{equation}
P_{\rm esc}=\frac{1}{2}\;\int^{\infty}_{-\infty}\Phi(x)\;E_{2}(\tau_{\rm d} + \tau \Phi(x))\;dx.
\label{eq:pescdust}
\end{equation}
This equation reflects the fact that dust grains provide a continuum opacity that is always capable of 
absorbing photons.
We have used this equation to calculate the exact escape probabilities in the presence of dust (Figure~\ref{fig:sprinkler}).
Because it is computationally burdensome to calculate the escape probability for all possible combinations of 
$\tau$ and $\tau_{\rm d}$, we seek an approximation such that the escape probability is the product of two 
separate, independent functions of $\tau$ and $\tau_{\rm d}$. 
On physical grounds, one expects this function to resemble 
$P_{\rm esc}(\tau,\tau_{\rm d}) \approx P_{\rm esc}(\tau_{\rm d}=0) \times E_{2}(\tau_{\rm d})$.
Clearly this is the appropriate expression when $\tau_{\rm d} \approx 0$.
It is also exact in the limit $\tau = 0$.
The deviation is expected to be strongest when $\tau \approx \tau_{\rm d} \approx 1$, but happily the 
deviation is not great.
Applying the slight modification, 
\begin{equation}
P_{\rm esc}^{\rm appr}(\tau,\tau_{\rm d}) = P_{\rm esc}(\tau_{\rm d}=0) \times E_{2}(0.95 \tau_{\rm d}),
\end{equation}
provides an excellent fit to the exact probability calculated using equation~\ref{eq:pescdust}, as illustrated 
in Figure~\ref{fig:sprinkler}, and tabulated in Table 1.
\begin{deluxetable}{lr}
\tablecolumns{2}
\small
\tablewidth{0pt}
\tablecaption{Maximum Error for $P_{\rm esc}$, with the inclusion of dust.}
\tablehead{\colhead{$\tau_d/\tau_0$} & \colhead{maximum error (\%)} }
\startdata
0.0&11.42\\
0.03&24.52\\
0.1&25.02\\
0.3&47.51\\
1.0&54.27\\
3.0&47.73\\
10.0&65.20\tablenotemark{a}\\
30.0&128.04\tablenotemark{a}\\
\enddata
\tablenotetext{a}{Maximum error was reached at high optical depth where the escape probability $<$ 10$^{-8}$.}
\end{deluxetable}
It can be seen that the approximations including dust fit very well in all regions except where $\tau \approx 1$
and $\tau_{\rm d} \approx 1$, as expected.
We can apply these escape probabilities to the calculation of the cooling rate from water when dust is present.

\subsection{Line Cooling}

The cooling per line in the optically thin limit is given by
\begin{equation}
\frac{\Lambda_{ul}}{n_{H_2O}}=S(T)\cdot\frac{8\pi kT}{\lambda^2}\left(\frac{hc/\lambda kT}{e^{hc/\lambda kT}-1}\right),
\label{eq:linecool}
\end{equation}
where $S(T)$ is the temperature-dependent line strength of the spectral line,
and $\lambda$ is the wavelength at line center (see Appendix).
In our calculations, we assume that level populations obey a Boltzmann distribution.  
This is justified if the number density of protoplanetary disks greatly exceeds the 
critical density of each transition.
As NK93 found no variations in level populations when the density exceeded 
$\approx 10^{10} \, {\rm cm}^{-3}$, we assume that all level populations have critical 
densities below this value.
Since we are motivated by chondrule formation, for which the gas densities exceed 
$\approx 10^{15} \, {\rm cm}^{-3}$, the use of a Boltzmann distribution is justified.
Unfortunately, for other problems (e.g., molecular clouds), our approach is limited in applicability.

The cooling rate is reduced below its optically thin limit as the column densities are increased,
due to the inability of photons to escape the system.
Specifically, the total cooling, $\mathcal{L}_{\rm LTE}$, is given by the total cooling per line, summed 
over all lines, including the escape probabilities:
\begin{equation}
\mathcal{L}_{\rm LTE}=\sum\Lambda_{ul}\;P_{\rm esc}(\tau_{ul},\tau_{\rm d}).
\label{eq16}
\end{equation}
Based on the methods of Plume et al. (2004), the gas optical depth at line center, $\tau_{0}$, can be found from the linestrength
and column density of water:
\begin{equation}
\tau_{0}=\frac{S(T)\:N_{H_2O}}{\left(\Delta v/c\right)\nu},
\label{eq12}
\end{equation}
(see Appendix), where the Doppler linewidth is given by
\begin{equation}
\Delta v = \left(\frac{2kT}{m_{H_2O}}\right)^{\frac{1}{2}},
\end{equation}
and is $\approx 1.36 \, {\rm km} \, {\rm s}^{-1}$ at $T = 2000 \, {\rm K}$.
In terms of the optical depth at line center, $\tau_{ul} = \tau_{0} \times \sqrt{\pi}$.

The dust optical depth at a depth $z$ into a semi-infinite volume is given by 
\begin{equation}
\tau_{\rm d} = \rho_{\rm g} \, z \, \kappa(\lambda),
\label{eq:taudust}
\end{equation}
where $\kappa(\lambda)$ is the opacity per gram of gas.
Assuming a dust-to-gas ratio $\rho_{\rm d} / \rho_{\rm g} = 5 \times 10^{-3}$, 
a particle radius $a_s = 0.5 \, \mu{\rm m}$, 
and an absorptivity $Q_{\rm abs} = 1$ for $\lambda < 2\pi a_s$ and $Q_{\rm abs} = 2\pi a_s/\lambda$ for $\lambda > 2\pi a_s$, 
we derive 
\begin{equation}
\kappa = 30 \, \min \left[ 1, (\lambda / 3.1 \, \mu{\rm m})^{-1} \right] \, {\rm cm}^{2} \, {\rm g}^{-1} 
\label{eq:ourkappa}
\end{equation}
We have determined the Rosseland mean dust opacity to be
$\kappa_{\rm R} = 28.76 \, {\rm cm}^{2} \, {\rm g}^{-1}$ at $T = 2000 \, {\rm K}$.
Our estimates are similar to those derived by Henning \& Stognienko (1996) using a particle size distribution instead of 
our simplified monodispersion.

\subsection{Cooling Rates}

\begin{figure}
\epsscale{0.8}    
\plotone{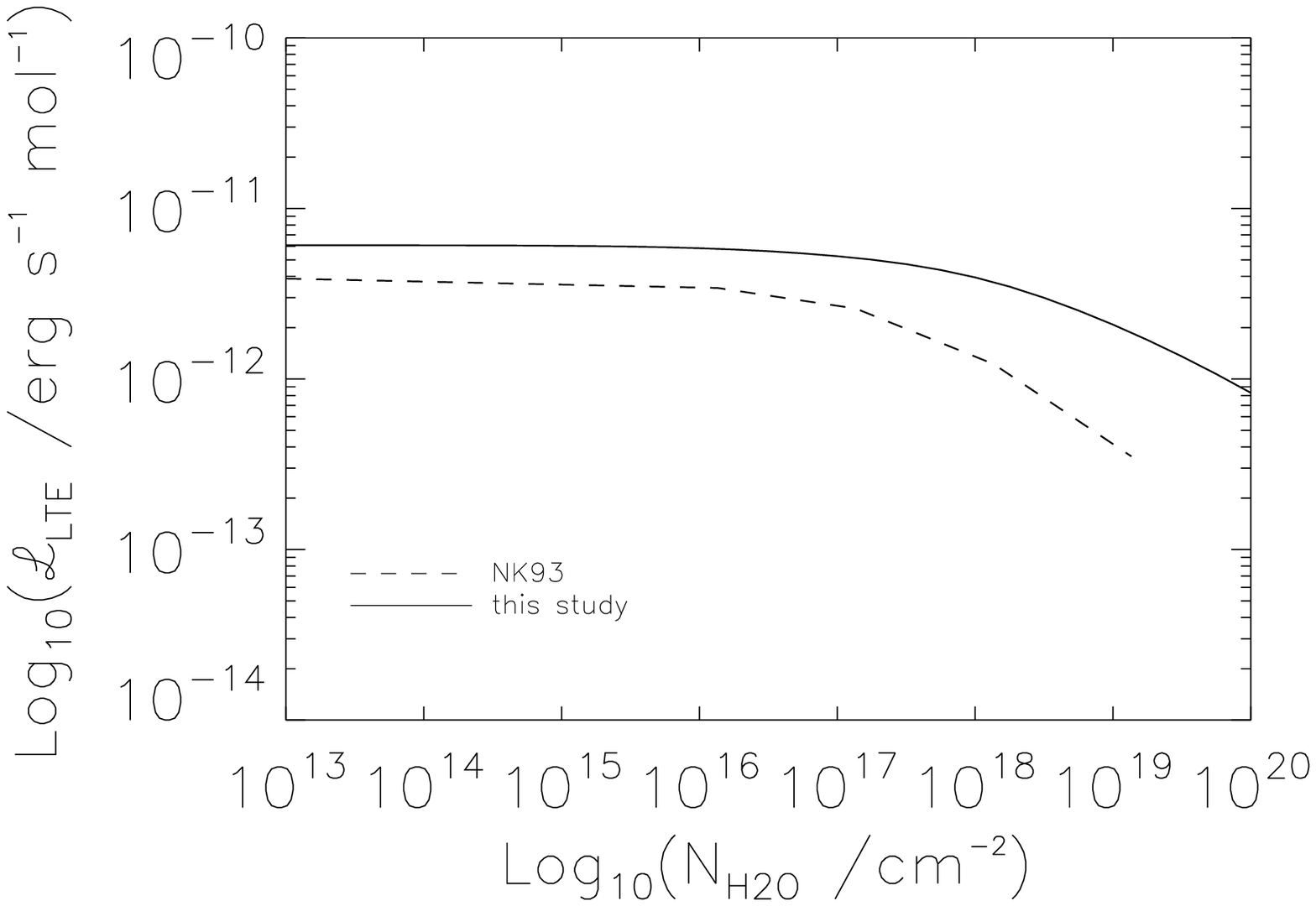}
\caption{The cooling rates of NK93 compared to those of this study.}
\label{fig:NK93comp}
\end{figure}

\begin{figure}
\epsscale{0.8}    
\plotone{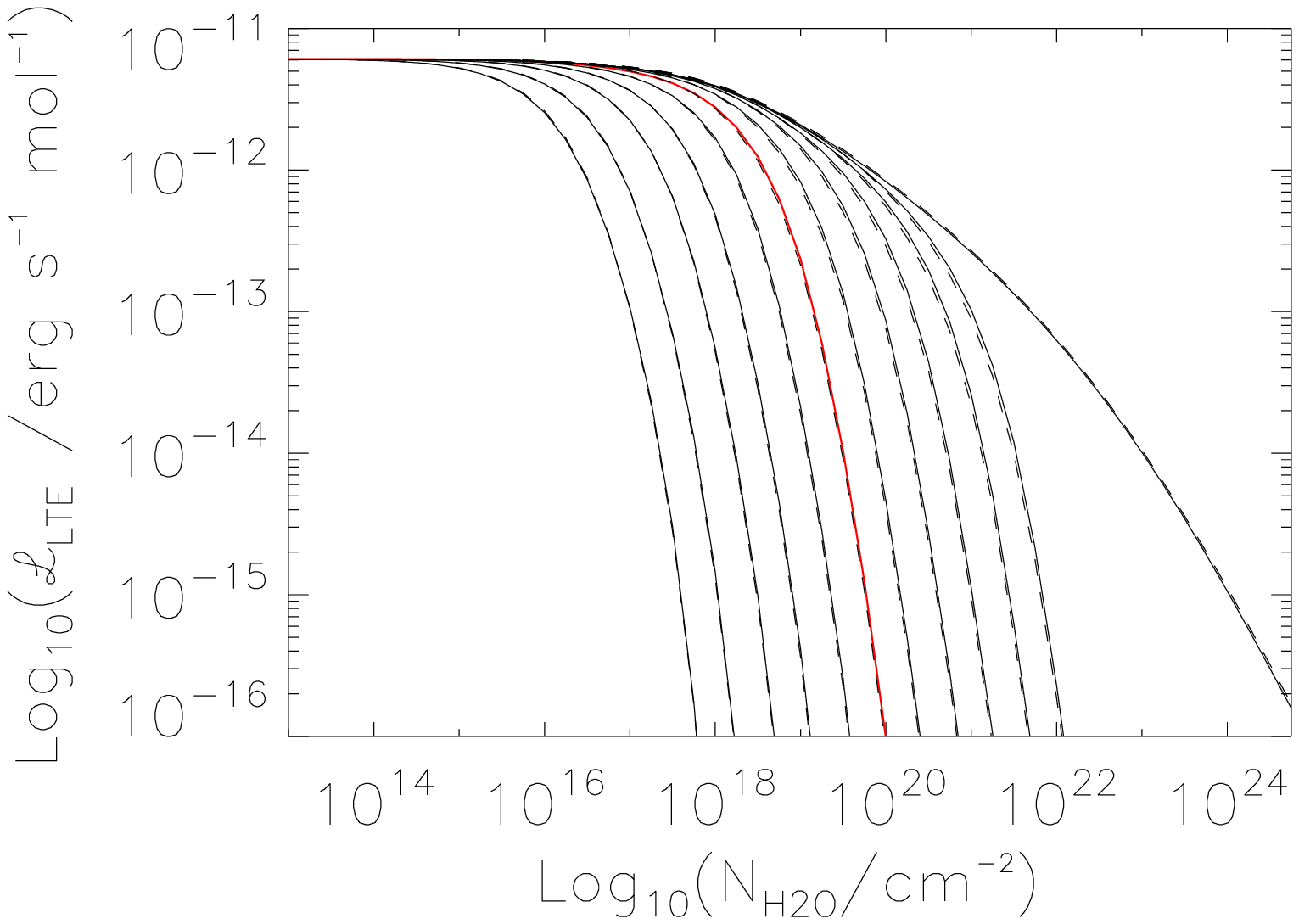}
\caption{Exact cooling rates due to $\water$ for various dust-to-water ratios (solid lines), as a 
function of water column density, and the cooling rates calculated using the approximation to the escape 
probability (dashed lines) for $T = 2000 \, {\rm K}$.
The rightmost curve is the case with no dust; the red curve is the canonical case (see text).  
The curves to the left of the canonical curve (from right to left) show cooling rates for 
3, 10, 30, 100, and 300 times the canonical dust-to-water ratio.
The curves to the right of the canonical curve (from left to right) are 1/3, 1/10, 1/30, 1/100, and 1/300 
times the canonical dust-to-water ratio.}
\label{fig:coolingrate}
\end{figure}

\begin{figure}
\epsscale{0.8}    
\plotone{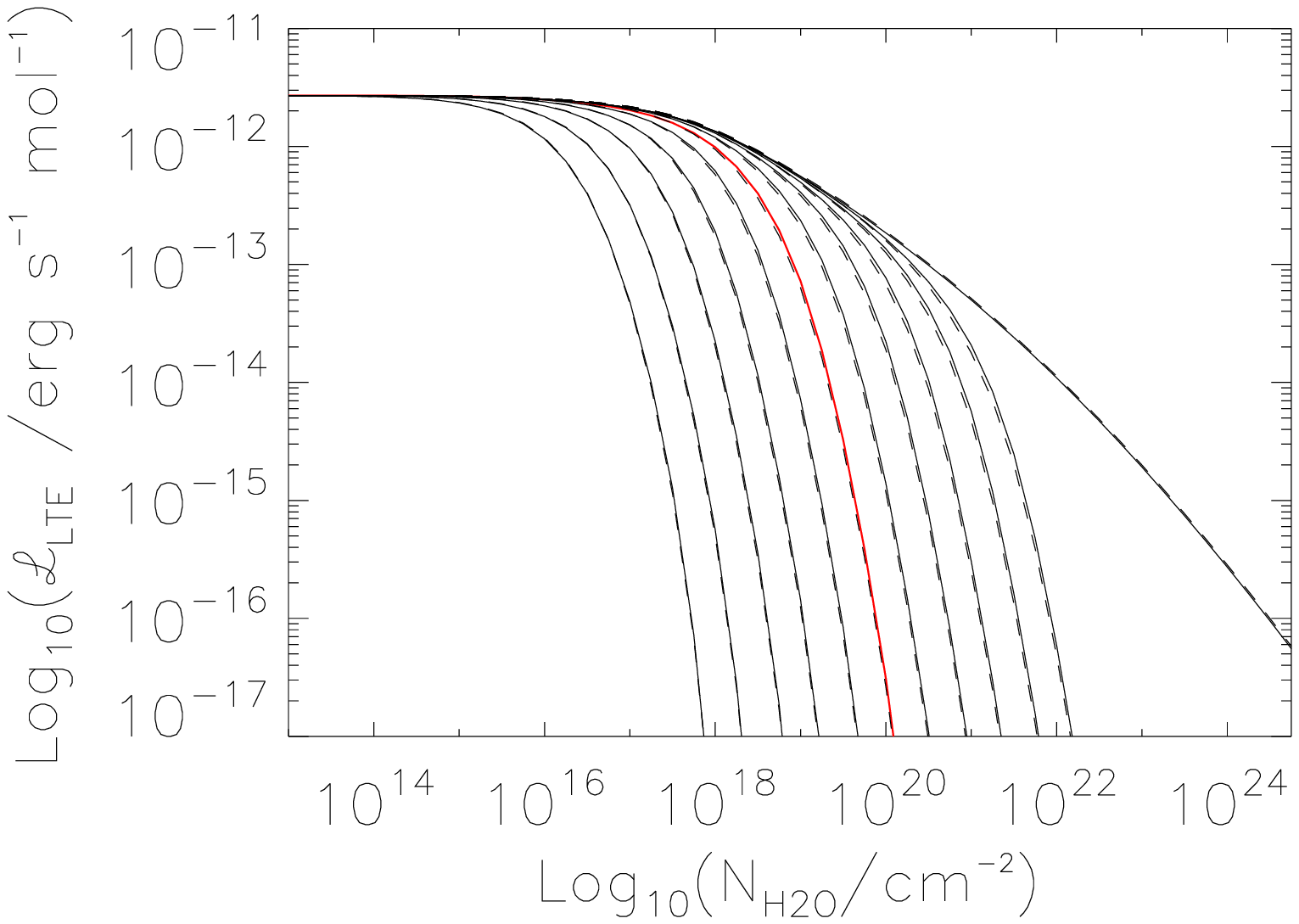}
\caption{Same as Figure~\ref{fig:coolingrate}, but with T~=~1500~K.}
\label{fig:coolrate1500}
\end{figure}

We first calculate the cooling rates in the absence of dust. 
Using many more transitions (and improved escape probabilities) we find that, in the absence of dust, our cooling rates 
are enhanced by about 30\% over those calculated by NK93 (Figure \ref{fig:NK93comp}).
For the cooling without dust, we found a slight difference ($< 6\%$) in cooling rate between the 
cases where the escape probabilities are calculated exactly, and when we use the approximation of 
Hollenbach \& McKee (1979). 
For consistency, we will quote cooling rates obtained using the 
Hollenbach \& McKee (1979) approximation.
The cooling rate for the $T = 2000 \, {\rm K}$ case, in the absence of dust, is plotted as the rightmost curve in Figure~\ref{fig:coolingrate}.
Also plotted in Figure~\ref{fig:coolingrate} are the cases where the dust-to-water ratio is varied from 300 times
its canonical value to 1/300 times its canonical value. 
Here the canonical case refers to a dust-to-gas mass ratio of 0.5\%, yielding the opacity discussed above; the 
dust-to-water ratio effectively measures only the dust opacity per gram of water vapor.
The canonical amount of water assumed here is a ratio ${\rm H}_{2}{\rm O} / {\rm H}_{2} = 8 \times 10^{-4}$.
A higher dust-to-water ratio means dust grains are more likely to absorb line photons, and the cooling rate
due to line emission is reduced below the optically thin limit at a smaller total column density of water.
Surprisingly, even small amounts of dust (300 times smaller than the canonical limit, or a dust-to-gas ratio
of $\sim 0.001 \%$) will prevent the gas from cooling as freely as it would without dust at column densities
$N_{\rm H2O} > 10^{22} \, {\rm cm}^{-2}$.  
The maximum error in each case between the exact value of the cooling rate (using the escape 
probability of equation 15) and that found using the approximation to the escape probability 
(equation 16) is shown in Table 2 (for $T = 2000 \, {\rm K}$).  
Figure~\ref{fig:coolrate1500} gives the same information for the case when $T = 1500 \, {\rm K}$.  
The cooling rates (assuming $a = 0$, the pure Doppler broadening case) for $T = 1250$ K, 1500 K, 1750 K, 
2000~K, and 2250 K, both in the absence of dust and with the canonical abundance of dust, are tabulated 
in Tables 3 through 12.   

\begin{deluxetable}{lr}
\tablecolumns{2}
\small
\tablewidth{0pt}
\tablecaption{Maximum Discrepancy between ``Exact" and Approximate Cooling Rates shown in 
Figure~\ref{fig:coolingrate}, at $T = 2000 \, {\rm K}$.}
\tablehead{\colhead{Dust-to-Water Ratio\tablenotemark{a}} & \colhead{maximum error (\%)} }
\startdata
0.0&5.54\\
1.0&14.01\\
3.0&13.29\\
10.0&12.24\\
30.0&11.04\\
100.0&9.26\\
300.0&7.45\\
1/3&14.77\\
1/10&15.64\\
1/30&15.39\\
1/100&17.93\\
1/300&18.40\\
\enddata
\tablenotetext{a}{Relative to the canonical value.}
\end{deluxetable} 

\begin{figure}
\epsscale{0.8}
\plotone{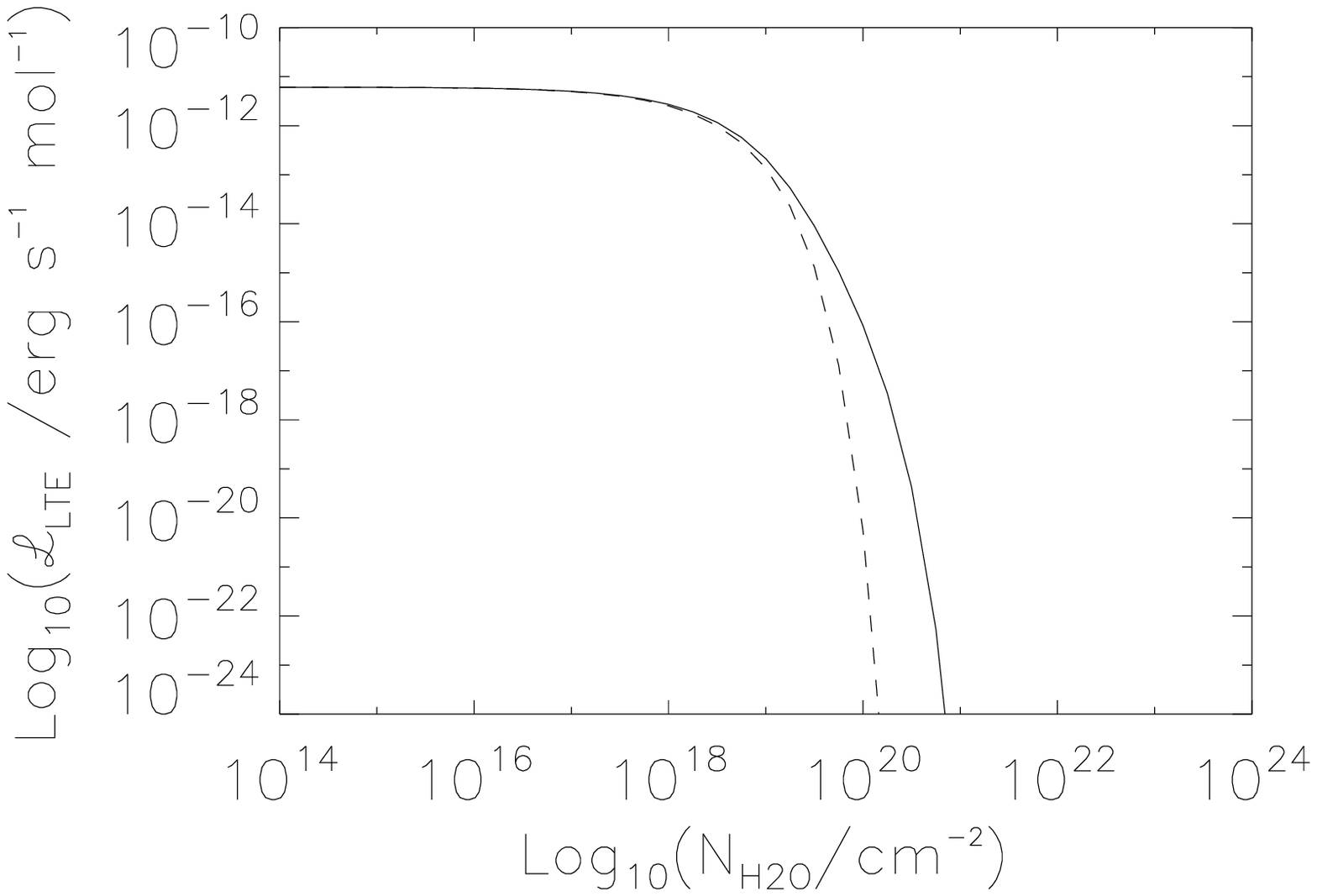}
\caption{Cooling rates due to $\water$ using the Rosseland mean dust opacity at
$T = 2000 \, {\rm K}$ (dashed line) and using the wavelength-dependent dust opacity (solid line).  
Most line cooling takes place at wavelengths longward of $3 \, \mu{\rm m}$, where the dust is
relatively optically thin.  Use of a single dust opacity at all wavelengths is not warranted.}
\label{fig:ross}
\end{figure}

Absorption of line photons by dust grains drastically affects the cooling of the gas.
The dust opacity is a strong function of wavelength, so we investigate the effects of assuming a constant opacity. 
In the continuum opacity case it would be reasonable to assume the flux of photons propagating through a gas
is equivalent to the flux derived assuming the gas had a wavelength-independent opacity equal to the 
Rosseland mean opacity. 
We therefore recalculated the cooling rate when the dust opacity equaled the Rosseland mean opacity cited above,
at all wavelengths. 
For the canonical dust-to-water ratio, the results are plotted in Figure~\ref{fig:ross}.
The optical depth of dust is given by
\begin{equation}
\tau_{\rm d} = \frac{ N_{\rm H2O} }{ n_{\rm H2O} / n_{\rm H2} } \, 1.4 m_{\rm H2} \, \kappa.
\end{equation}
Using the Rosseland mean opacity and canonical water-to-gas ratios cited above, we would estimate 
$\tau \approx 1$ when $N_{\rm H2O} \approx 6 \times 10^{18} \, {\rm cm}^{-2}$.
This is indeed the point where the cooling rate is significantly reduced due to absorption by dust grains.
When the opacity is replaced by the wavelength-dependent opacity $\kappa(\lambda)$, however, the cooling
is not effectively reduced until higher column densities are reached.  
This signifies that most of the cooling is effected by emission of line photons with wavelengths at 
which $\kappa(\lambda) < \kappa_{\rm R}$, i.e., $\lambda > 3 \, \mu{\rm m}$.
Use of a wavelength-dependent opacity in conjunction with a calculation of the cooling at each ${\rm H}_{2}{\rm O}$
wavelength is therefore necessary.
We use the wavelength-dependent opacity in all cases cited here.  

\begin{figure}
\epsscale{0.8}        
\plotone{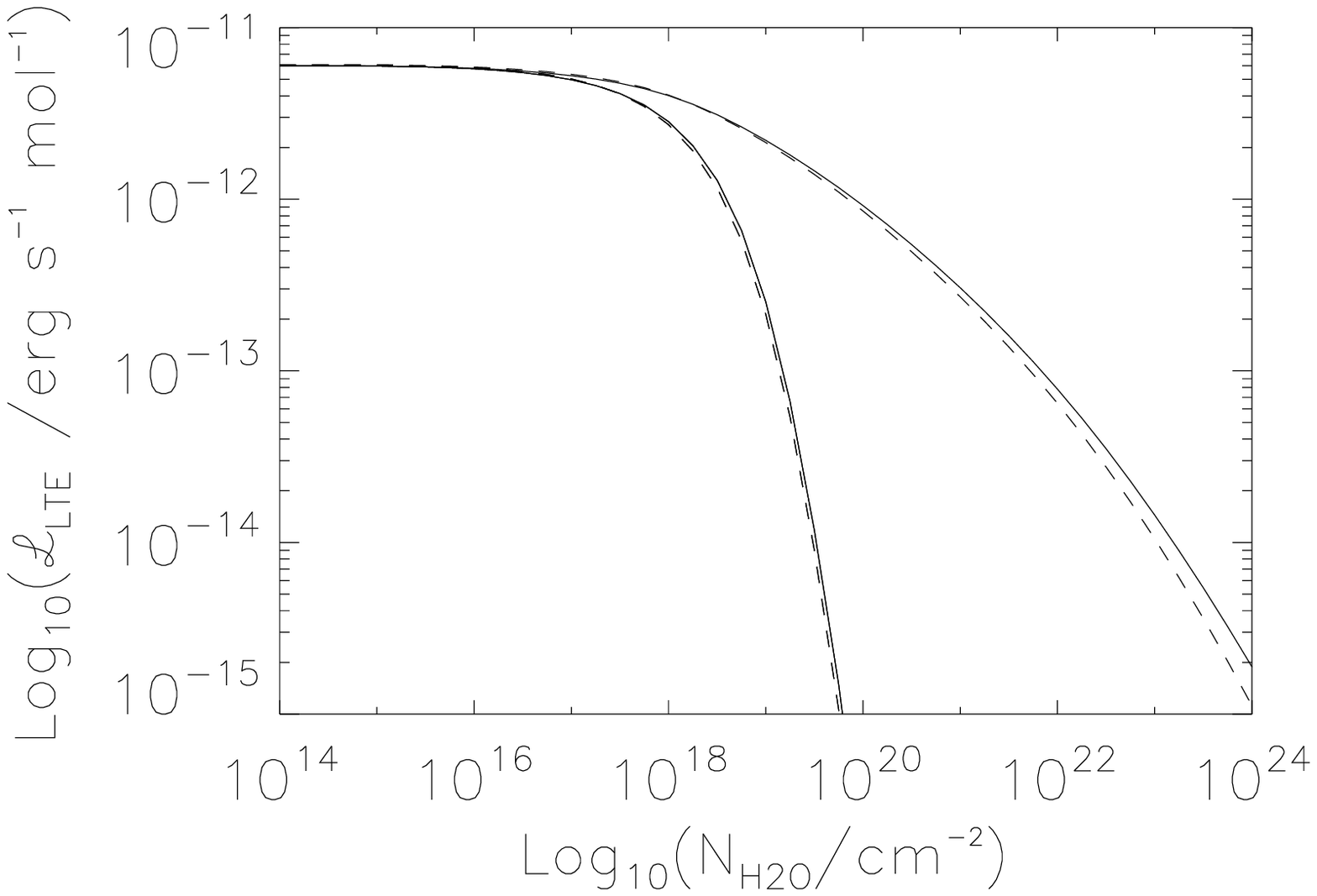}
\caption{Cooling rates with a broadened line profile corresponding to $a=0.1$ (solid curves), 
compared to those calculated assuming pure Doppler broadening $a = 0$ (dashed curves), both 
in the absence of dust (top curves) and with a canonical dust-to-water ratio (bottom curves). }
\label{fig:acool}
\end{figure}

Finally, we investigate the effect of the line broadening parameter $a$ on the cooling rate. 
Increasing $a$ above the $a=0$ pure Doppler broadening case has the effect of putting more emission in the 
optically thin wings of the line profile.  
Effectively, this should act like a reduction of the overall water column density.
We have calculated the cooling rates assuming line profiles with $a = 0$  and $a = 0.1$,  
in the cases where no dust is present, and when it is present at the canonical value. 
These cooling rates are plotted in Figure~\ref{fig:acool}. 
As expected, the cooling rate is slightly higher when $a = 0.1$, because the column density of
water is effectively reduced; however, the difference is slight. 
In the case with no dust, the discrepancy between the $a = 0$ and $a = 0.1$ cases is $< 11\%$ for 
a column density of water of $N_{\rm H2O} < 10^{21} \, {\rm cm}^{-2}$, increasing to $\approx 41\%$ at 
a column density of $10^{24} \, {\rm cm}^{-2}$.  
In the case with dust, the discrepancy is $< 13\%$ at $N_{\rm H2O} < 10^{19} \, {\rm cm}^{-2}$.
These calculations are much more computationally burdensome than the $a=0$ pure Doppler broadening case, 
because of the need to calculate the Voigt function. 
Fortunately, it is seen that their effects on the cooling rates of gas are minimal, $< 10\%$ in the regime
where cooling is significant, for $a$ up to 0.1.
As $a$ is typically $\ll 1$ (Mihalas 1978; Bowers \& Deeming 1984), we will assume $a = 0$ in the cases that follow.

\begin{deluxetable}{lccr}
\tablecolumns{4}
\small
\tablewidth{0pt}
\tablecaption{Cooling rates without the inclusion of dust, $a=0.0$, $T = 1250 \, {\rm K}$}
\tablehead{\colhead{$N_{H_2O}$} & \colhead{$\mathcal{L}_{\rm LTE_{exact}}$} & \colhead{$\mathcal{L}_{\rm LTE_{approx}}$} & \colhead{\% error} }
\startdata
10$^{13}$&1.547 x 10$^{-12}$&1.548 x 10$^{-12}$&0.03\\
10$^{14}$&1.543 x 10$^{-12}$&1.546 x 10$^{-12}$&0.16\\
10$^{15}$&1.517 x 10$^{-12}$&1.526 x 10$^{-12}$&0.61\\
10$^{16}$&1.405 x 10$^{-12}$&1.425 x 10$^{-12}$&1.44\\
10$^{17}$&1.128 x 10$^{-12}$&1.160 x 10$^{-12}$&2.82\\
10$^{18}$&6.259 x 10$^{-13}$&6.415 x 10$^{-13}$&2.95\\
10$^{19}$&2.227 x 10$^{-13}$&2.293 x 10$^{-13}$&2.96\\
10$^{20}$&6.611 x 10$^{-14}$&6.816 x 10$^{-14}$&3.10\\
10$^{21}$&1.615 x 10$^{-14}$&1.668 x 10$^{-14}$&3.33\\
10$^{22}$&3.318 x 10$^{-14}$&3.437 x 10$^{-14}$&3.58\\
10$^{23}$&5.750 x 10$^{-16}$&5.978 x 10$^{-16}$&3.95\\
10$^{24}$&8.587 x 10$^{-17}$&8.959 x 10$^{-17}$&4.34\\
\enddata
\end{deluxetable} 
%

\begin{deluxetable}{lccr}
\tablecolumns{4}
\small
\tablewidth{0pt}
\tablecaption{Cooling rates with the canonical dust-to-water ratio (see text), $a=0.0$, $T = 1250 \, {\rm K}$}
\tablehead{\colhead{$N_{H_2O}$} & \colhead{$\mathcal{L}_{\rm LTE_{exact}}$} & \colhead{$\mathcal{L}_{\rm LTE_{approx}}$} & \colhead{\% error} }
\startdata
10$^{13}$&1.547 x 10$^{-12}$&1.548 x 10$^{-12}$&0.03\\
10$^{14}$&1.543 x 10$^{-12}$&1.546 x 10$^{-12}$&0.16\\
10$^{15}$&1.516 x 10$^{-12}$&1.525 x 10$^{-12}$&0.59\\
10$^{16}$&1.397 x 10$^{-12}$&1.415 x 10$^{-12}$&1.26\\
10$^{17}$&1.083 x 10$^{-12}$&1.097 x 10$^{-12}$&1.62\\
10$^{18}$&4.686 x 10$^{-13}$&4.439 x 10$^{-13}$&5.27\\
10$^{19}$&3.204 x 10$^{-14}$&2.765 x 10$^{-14}$&13.71\\
10$^{20}$&1.459 x 10$^{-17}$&1.226 x 10$^{-17}$&15.98\\
10$^{21}$&1.101 x 10$^{-28}$&2.148 x 10$^{-28}$&95.04\\
10$^{22}$&$0.00$&$0.00$&0.00\\
10$^{23}$&$0.00$&$0.00$&0.00\\
10$^{24}$&$0.00$&$0.00$&0.00\\
\enddata
\end{deluxetable} 
%

\begin{deluxetable}{lccr}
\tablecolumns{4}
\small
\tablewidth{0pt}
\tablecaption{Cooling rates without the inclusion of dust, $a=0.0$, $T = 1500 \, {\rm K}$}
\tablehead{\colhead{$N_{H_2O}$} & \colhead{$\mathcal{L}_{\rm LTE_{exact}}$} & \colhead{$\mathcal{L}_{\rm LTE_{approx}}$} & \colhead{\% error} }
\startdata
10$^{13}$&2.704 x 10$^{-12}$&2.705 x 10$^{-12}$&0.02\\
10$^{14}$&2.700 x 10$^{-12}$&2.704 x 10$^{-12}$&0.12\\
10$^{15}$&2.667 x 10$^{-12}$&2.679 x 10$^{-12}$&0.47\\
10$^{16}$&2.519 x 10$^{-12}$&2.548 x 10$^{-12}$&1.15\\
10$^{17}$&2.127 x 10$^{-12}$&2.179 x 10$^{-12}$&2.47\\
10$^{18}$&1.347 x 10$^{-12}$&1.383 x 10$^{-12}$&2.81\\
10$^{19}$&5.597 x 10$^{-13}$&5.755 x 10$^{-13}$&2.84\\
10$^{20}$&1.850 x 10$^{-13}$&1.905 x 10$^{-13}$&3.00\\
10$^{21}$&4.992 x 10$^{-14}$&5.153 x 10$^{-14}$&3.26\\
10$^{22}$&1.094 x 10$^{-14}$&1.132 x 10$^{-14}$&3.45\\
10$^{23}$&1.917 x 10$^{-15}$&1.991 x 10$^{-15}$&3.86\\
10$^{24}$&2.715 x 10$^{-16}$&2.837 x 10$^{-16}$&4.48\\
\enddata
\end{deluxetable} 
%

\begin{deluxetable}{lccr}
\tablecolumns{4}
\small
\tablewidth{0pt}
\tablecaption{Cooling rates with the canonical dust-to-water ratio (see text), $a=0.0$, $T = 1500 \, {\rm K}$}
\tablehead{\colhead{$N_{H_2O}$} & \colhead{$\mathcal{L}_{\rm LTE_{exact}}$} & \colhead{$\mathcal{L}_{\rm LTE_{approx}}$} & \colhead{\% error} }
\startdata
10$^{13}$&2.704 x 10$^{-12}$&2.705 x 10$^{-12}$&0.02\\
10$^{14}$&2.699 x 10$^{-12}$&2.702 x 10$^{-12}$&0.12\\
10$^{15}$&2.664 x 10$^{-12}$&2.677 x 10$^{-12}$&0.46\\
10$^{16}$&2.502 x 10$^{-12}$&2.527 x 10$^{-12}$&1.01\\
10$^{17}$&2.031 x 10$^{-12}$&2.055 x 10$^{-12}$&1.36\\
10$^{18}$&9.849 x 10$^{-13}$&9.430 x 10$^{-13}$&4.26\\
10$^{19}$&7.223 x 10$^{-13}$&6.338 x 10$^{-13}$&12.26\\
10$^{20}$&3.081 x 10$^{-17}$&2.612 x 10$^{-17}$&15.23\\
10$^{21}$&2.210 x 10$^{-28}$&4.353 x 10$^{-28}$&97.01\\
10$^{22}$&$0.00$&$0.00$&0.00\\
10$^{23}$&$0.00$&$0.00$&0.00\\
10$^{24}$&$0.00$&$0.00$&0.00\\
\enddata
\end{deluxetable} 
%

\begin{deluxetable}{lccr}
\tablecolumns{4}
\small
\tablewidth{0pt}
\tablecaption{Cooling rates without the inclusion of dust, $a=0.0$, $T = 1750 \, {\rm K}$}
\tablehead{\colhead{$N_{H_2O}$} & \colhead{$\mathcal{L}_{\rm LTE_{exact}}$} & \colhead{$\mathcal{L}_{\rm LTE_{approx}}$} & \colhead{\% error} }
\startdata
10$^{13}$&4.219 x 10$^{-12}$&4.220 x 10$^{-12}$&0.01\\
10$^{14}$&4.214 x 10$^{-12}$&4.217 x 10$^{-12}$&0.08\\
10$^{15}$&4.177 x 10$^{-12}$&4.193 x 10$^{-12}$&0.37\\
10$^{16}$&4.002 x 10$^{-12}$&4.040 x 10$^{-12}$&0.94\\
10$^{17}$&3.501 x 10$^{-12}$&3.575 x 10$^{-12}$&2.11\\
10$^{18}$&2.446 x 10$^{-12}$&2.511 x 10$^{-12}$&2.68\\
10$^{19}$&1.158 x 10$^{-12}$&1.189 x 10$^{-12}$&2.74\\
10$^{20}$&4.219 x 10$^{-13}$&4.343 x 10$^{-13}$&2.93\\
10$^{21}$&1.240 x 10$^{-13}$&1.279 x 10$^{-13}$&3.15\\
10$^{22}$&2.863 x 10$^{-14}$&2.958 x 10$^{-14}$&3.35\\
10$^{23}$&4.913 x 10$^{-15}$&5.101 x 10$^{-15}$&3.83\\
10$^{24}$&6.215 x 10$^{-16}$&1.156 x 10$^{-16}$&5.75\\
\enddata
\end{deluxetable} 
%

\begin{deluxetable}{lccr}
\tablecolumns{4}
\small
\tablewidth{0pt}
\tablecaption{Cooling rates for the canonical dust-to-water ratio (see text), $a=0.0$, $T = 1750 \, {\rm K}$}
\tablehead{\colhead{$N_{H_2O}$} & \colhead{$\mathcal{L}_{\rm LTE_{exact}}$} & \colhead{$\mathcal{L}_{\rm LTE_{approx}}$} & \colhead{\% error} }
\startdata
10$^{13}$&4.219 x 10$^{-12}$&4.220 x 10$^{-12}$&0.01\\
10$^{14}$&4.213 x 10$^{-12}$&4.217 x 10$^{-12}$&0.08\\
10$^{15}$&4.173 x 10$^{-12}$&4.188 x 10$^{-12}$&0.35\\
10$^{16}$&3.973 x 10$^{-12}$&4.006 x 10$^{-12}$&0.82\\
10$^{17}$&3.329 x 10$^{-12}$&3.365 x 10$^{-12}$&1.15\\
10$^{18}$&1.754 x 10$^{-12}$&1.696 x 10$^{-12}$&3.27\\
10$^{19}$&1.383 x 10$^{-12}$&1.233 x 10$^{-12}$&10.84\\
10$^{20}$&5.735 x 10$^{-17}$&4.899 x 10$^{-17}$&14.57\\
10$^{21}$&3.973 x 10$^{-28}$&7.831 x 10$^{-28}$&97.13\\
10$^{22}$&$0.00$&$0.00$&0.00\\
10$^{23}$&$0.00$&$0.00$&0.00\\
10$^{24}$&$0.00$&$0.00$&0.00\\
\enddata
\end{deluxetable} 
%

\begin{deluxetable}{lccr}
\tablecolumns{4}
\small
\tablewidth{0pt}
\tablecaption{Cooling rates without the inclusion of dust, $a=0.0$, $T = 2000 \, {\rm K}$}
\tablehead{\colhead{$N_{H_2O}$} & \colhead{$\mathcal{L}_{\rm LTE_{exact}}$} & \colhead{$\mathcal{L}_{\rm LTE_{approx}}$} & \colhead{\% error} }
\startdata
10$^{13}$&6.083 x 10$^{-12}$&6.084 x 10$^{-12}$&0.01\\
10$^{14}$&6.078 x 10$^{-12}$&6.082 x 10$^{-12}$&0.02\\
10$^{15}$&6.040 x 10$^{-12}$&6.057 x 10$^{-12}$&0.28\\
10$^{16}$&5.847 x 10$^{-12}$&5.893 x 10$^{-12}$&0.78\\
10$^{17}$&5.250 x 10$^{-12}$&5.344 x 10$^{-12}$&1.79\\
10$^{18}$&3.943 x 10$^{-12}$&4.045 x 10$^{-12}$&2.58\\
10$^{19}$&2.084 x 10$^{-12}$&2.139 x 10$^{-12}$&2.68\\
10$^{20}$&8.297 x 10$^{-13}$&8.536 x 10$^{-13}$&2.89\\
10$^{21}$&2.617 x 10$^{-13}$&2.698 x 10$^{-13}$&3.12\\
10$^{22}$&6.260 x 10$^{-14}$&6.464 x 10$^{-14}$&3.26\\
10$^{23}$&1.013 x 10$^{-14}$&1.724 x 10$^{-14}$&3.83\\
10$^{24}$&1.060 x 10$^{-14}$&1.118 x 10$^{-14}$&5.54\\
\enddata
\end{deluxetable} 
%

\begin{deluxetable}{lccr}
\tablecolumns{4}
\small
\tablewidth{0pt}
\tablecaption{Cooling rates for the canonical dust-to-water ratio (see text), $a=0.0$, $T = 2000 \, {\rm K}$}
\tablehead{\colhead{$N_{H_2O}$} & \colhead{$\mathcal{L}_{\rm LTE_{exact}}$} & \colhead{$\mathcal{L}_{\rm LTE_{approx}}$} & \colhead{\% error} }
\startdata
10$^{13}$&6.083 x 10$^{-12}$&6.084 x 10$^{-12}$&0.01\\
10$^{14}$&6.077 x 10$^{-12}$&6.081 x 10$^{-12}$&0.06\\
10$^{15}$&6.034 x 10$^{-12}$&6.051 x 10$^{-12}$&0.28\\
10$^{16}$&5.803 x 10$^{-12}$&5.842 x 10$^{-12}$&0.69\\
10$^{17}$&4.977 x 10$^{-12}$&5.025 x 10$^{-12}$&0.98\\
10$^{18}$&2.783 x 10$^{-12}$&2.716 x 10$^{-12}$&2.39\\
10$^{19}$&2.351 x 10$^{-12}$&2.129 x 10$^{-12}$&9.45\\
10$^{20}$&9.721 x 10$^{-17}$&8.359 x 10$^{-17}$&14.01\\
10$^{21}$&6.497 x 10$^{-28}$&1.283 x 10$^{-27}$&97.59\\
10$^{22}$&$0.00$&$0.00$&0.00\\
10$^{23}$&$0.00$&$0.00$&0.00\\
10$^{24}$&$0.00$&$0.00$&0.00\\
\enddata
\end{deluxetable} 
%
\begin{deluxetable}{lccr}
\tablecolumns{4}
\small
\tablewidth{0pt}
\tablecaption{Cooling rates without the inclusion of dust, $a=0.0$, $T = 2250\, {\rm K}$}
\tablehead{\colhead{$N_{H_2O}$} & \colhead{$\mathcal{L}_{\rm LTE_{exact}}$} & \colhead{$\mathcal{L}_{\rm LTE_{approx}}$} & \colhead{\% error} }
\startdata
10$^{13}$&8.284 x 10$^{-12}$&8.285 x 10$^{-12}$&0.01\\
10$^{14}$&8.279 x 10$^{-12}$&8.283 x 10$^{-12}$&0.04\\
10$^{15}$&8.241 x 10$^{-12}$&8.260 x 10$^{-12}$&0.22\\
10$^{16}$&8.040 x 10$^{-12}$&8.092 x 10$^{-12}$&0.65\\
10$^{17}$&7.364 x 10$^{-12}$&7.476 x 10$^{-12}$&1.52\\
10$^{18}$&5.835 x 10$^{-12}$&5.979 x 10$^{-12}$&2.47\\
10$^{19}$&3.383 x 10$^{-12}$&3.471 x 10$^{-12}$&2.63\\
10$^{20}$&1.457 x 10$^{-12}$&1.499 x 10$^{-12}$&2.85\\
10$^{21}$&4.870 x 10$^{-13}$&5.022 x 10$^{-13}$&3.11\\
10$^{22}$&1.185 x 10$^{-13}$&1.222 x 10$^{-13}$&3.20\\
10$^{23}$&1.726 x 10$^{-14}$&1.793 x 10$^{-14}$&3.87\\
10$^{24}$&1.440 x 10$^{-15}$&1.543 x 10$^{-15}$&7.14\\
\enddata
\end{deluxetable} 
%

\begin{deluxetable}{lccr}
\tablecolumns{4}
\small
\tablewidth{0pt}
\tablecaption{Cooling rates for the canonical dust-to-water ratio (see text), $a=0.0$, $T = 2250 \, {\rm K}$}
\tablehead{\colhead{$N_{H_2O}$} & \colhead{$\mathcal{L}_{\rm LTE_{exact}}$} & \colhead{$\mathcal{L}_{\rm LTE_{approx}}$} & \colhead{\% error} }
\startdata
10$^{13}$&8.284 x 10$^{-12}$&8.285 x 10$^{-12}$&0.01\\
10$^{14}$&8.279 x 10$^{-12}$&8.281 x 10$^{-12}$&0.04\\
10$^{15}$&8.233 x 10$^{-12}$&8.250 x 10$^{-12}$&0.22\\
10$^{16}$&7.976 x 10$^{-12}$&8.022 x 10$^{-12}$&0.58\\
10$^{17}$&6.963 x 10$^{-12}$&7.023 x 10$^{-12}$&0.86\\
10$^{18}$&4.065 x 10$^{-12}$&3.998 x 10$^{-12}$&1.64\\
10$^{19}$&3.658 x 10$^{-12}$&3.361 x 10$^{-12}$&8.11\\
10$^{20}$&1.532 x 10$^{-16}$&1.324 x 10$^{-16}$&13.56\\
10$^{21}$&9.895 x 10$^{-28}$&1.954 x 10$^{-27}$&97.46\\
10$^{22}$&$0.00$&$0.00$&0.00\\
10$^{23}$&$0.00$&$0.00$&0.00\\
10$^{24}$&$0.00$&$0.00$&0.00\\
\enddata
\end{deluxetable} 

\section{Cooling During Chondrule-forming Shocks}

Line cooling is effective at cooling gas, provided the line photons are not reabsorbed by $\water$ molecules or dust grains before they can escape to the cool, pre-shock region.  In a nebular shock, it is assured that sufficiently far from the shock front, no line photons can escape and the gas and dust will become thermally coupled and cool slowly, if at all.  What is not clear is the degree to which line cooling is significant in the region immediately past the shock.  Does line cooling
lead to a significant drop in temperature ($> 10^2$ K) before
dust grains begin to reabsorb line photons?  Are the cooling
rates of chondrules dominated by line cooling and therefore
high ($\sim 10^4 \, {\rm K} \, {\rm hr}^{-1}$) at the temperatures 
at which chondrules crystallize, as found by INSN, or can line
cooling be neglected, as DC02 and CH02 assume?  To answer
these questions we have constructed a toy model to assess 
the maximum possible importance of line cooling and to 
determine whether more detailed calculations of chondrule
formation in nebular shocks need to include line cooling.  

Our toy model builds on the calculation presented in Section 1.  We find it convenient to convert a time derivative $\partial / \partial t$ to a 
spatial derivative $V_{\rm g} \partial / \partial z$, and then convert the spatial variable $z$
(distance past the shock front) into a column density of water past the shock front (assuming the 
water density remains constant).
Then the cooling with column density of water is given by 
\begin{equation}
\frac{\partial T}{\partial N_{H_20}} = -\left( \frac{1}{\rho_{\rm g} V_{\rm g} } \right) \, 
\left( \frac{m_{\rm H}}{k} \right) \, \mathcal{L}_{\rm LTE}\left(N_{H2O}\right).
\label{eq16.1}
\end{equation}
The cooling rate is a function of the water column density but also the dust-to-water ratio
(via ${\cal L}_{\rm LTE}$).
Given a dust-to-water ratio, this equation can then be integrated to find $T$ as a function of 
$N_{\rm H2O}$.
Note that we have not accounted for the fact that the line broadening should decrease as 
the temperature drops, making it slightly harder for the gas to cool; our simplified analysis
therefore overestimates the cooling somewhat. 

\begin{figure}
\epsscale{0.8}      
\plotone{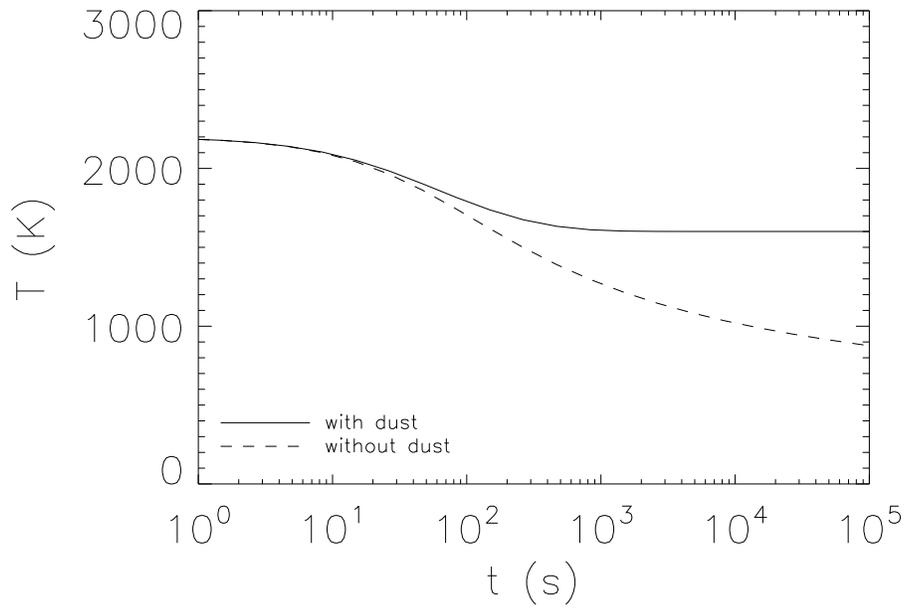}
\caption{Gas temperature as a function of time, both with and without the inclusion of dust grains that
 can absorb line photons (see text for details).}
\label{fig:chondrule}
\end{figure}

We plot the results of this integration in Figure~\ref{fig:chondrule} for both the dust-free case and the case where the dust is one-tenth that of the canonical value (in order to show the effect of even a small amount of dust).
For ease of comparison to chondrules we have also assumed the canonical water-to-gas ratio 
(and other canonical values) to convert $N_{\rm H2O}$ to a time. 
Specifically, we assumed a pre-shock density of $\rho_{\rm g} = 10^{-9} \, {\rm g} \, {\rm cm}^{-3}$, 
a shock velocity of $V_{\rm s} = 7 \, {\rm km} \, {\rm s}^{-1}$, assumed the density increased and the
velocity decreased by a factor of $(\gamma + 1) / (\gamma - 1) = 6$ past the shock front, 
and used our canonical water-to-gas ratio of $n_{H2O} / n_{H2} = 8 \times 10^{-4}$.
These parameters are consistent with our assumption of an initial post-shock temperature of 2200 K 
(DC02; CH02).
In the absence of dust, the gas would cool below 1400 K in roughly 300 seconds, leading to a cooling 
rate of $> 1 \times 10^{4} \, {\rm K} \, {\rm hr}^{-1}$.
This cooling is attributable solely to effective cooling by water; by comparison to NK93 we infer
these are mostly vibrational photons.
As chondrules and gas are expected to be thermally well coupled more than 100 seconds past the 
shock front (DC02; CH02), this can be interpreted as a likely cooling rate of chondrules as well.
The cooling rate begins to taper off as large column densities are reached, but not until times 
$> 10^{5} \, {\rm s}$, by which time the gas and chondrules have traveled roughly $10^{10} \, {\rm cm}$
past the shock front, equivalent to a column density of water $> 10^{22} \, {\rm cm}^{-2}$.

Including the absorption of line photons by dust grains completely changes the thermal history of 
the gas and chondrules. 
Using an opacity of dust $\sim 30 \, {\rm cm}^{2} \, {\rm g}^{-1}$ (at short wavelengths) and a post-shock 
gas density of 
$6 \times 10^{-9} \, {\rm g} \, {\rm cm}^{-3}$, the optical depth of dust exceeds unity after only
55 km past the shock front, which is reached after traveling only 48 seconds.  
During this first minute or so past the shock front, the line photons emitted by ${\rm H}_{2}{\rm O}$
molecules escape freely and effectively cool the gas; but after the first minute or so, they are 
absorbed by dust grains instead.
Instead of $\water$ line photons freely escaping to the shock front and cooling the gas and dust system, the photons are reabsorbed ``on-the-spot'' and do not cool the gas. 
This effect of the reduced cooling rate is clearly seen in Figure~\ref{fig:chondrule} beyond about
$10^2$ seconds.
The trapping of line photons by dust grains becomes so effective, no further cooling by line emission
is possible after a few minutes.
Ignoring the effects of cooling other than line photons, the temperature would stabilize at about 
1600 K.
For this particular choice of parameters, chondrules would {\it not} cool at rates 
$\sim 10^4 \, {\rm K} \, {\rm hr}^{-1}$ through their crystallization temperatures (1400 - 1800 K)
as in INSN and Miura \& Nakamoto (2006);
in fact, they wouldn't cool at all under the assumptions of our toy model.

We have investigated the effect of varying the dust-to-gas and water-to-gas ratios on the cooling rates of chondrules.  
Their behaviors are similar to those depicted in Figure~\ref{fig:chondrule}.
In Table 4 we report only the average cooling rates of chondrules over their crystallization temperature range (i.e., 
from 1800 K to 1400 K).
The gas density and shock velocity are not varied. 
The results of this parameter study show that even a small amount of dust very effectively shuts down the line cooling 
due to \wwater.

\begin{deluxetable}{cccc}
\tablecolumns{4}
\small
\tablewidth{0pt}
\tablecaption{Cooling Rates for various dust to gas and water to gas ratios}
\tablehead{\colhead{dust/gas\tablenotemark{a}} & \colhead{water/gas\tablenotemark{b}} & \colhead{$T$ at 10$^4$ s (K)} & \colhead{Cooling Rate \tablenotemark{c}} }
\startdata
0.0  & 0.1  & 1275 & $2.7 \times 10^{2}$ \\
0.0  & 1.0  & 1000 & $3.2 \times 10^{3}$ \\
0.0  & 10.0 &  875 & $3.7 \times 10^{4}$ \\
0.1  & 0.1  & 2000 & N/A \\
0.1  & 1.0  & 1650 &\tablenotemark{d}\\
0.1  & 10.0 & 1225 & $2.3 \times 10^{4}$ \\
1.0  & 0.1  & 2175 & N/A             \\
1.0  & 1.0  & 2000 & N/A             \\
1.0  & 10.0 & 1600 &\tablenotemark{e}\\
10.0 & 0.1  & 2200 & N/A             \\
10.0 & 1.0  & 2175 & N/A             \\
10.0 & 10.0 & 2000 & N/A             \\
\enddata
\tablenotetext{a}{times the canonical dust/gas mass ratio $5 \times 10^{-3}$}
\tablenotetext{b}{times the canonical water/gas ratio $n_{\rm H2O} / n_{\rm H2} = 8 \times 10^{-4}$}
\tablenotetext{c}{the average cooling rate between 1800 and 1400 K, in ${\rm K} \, {\rm hr}^{-1}$}
\tablenotetext{d}{$490 \, {\rm K} \, {\rm hr}^{-1}$ between 1800 and 1650 K; never cooled below 1650 K}
\tablenotetext{e}{$5000\, {\rm K} \, {\rm hr}^{-1}$ between 1800 and 1600 K; never cooled below 1600 K}
\end{deluxetable} 

\section{Discussion and Conclusions}

We have built upon the line cooling work of NK93, by using a much expanded database of spectral lines, improved escape probabilities, and the inclusion of dust grains.  We have found that although the cooling rate due to $\water$ line cooling is a complicated combination of gas and dust opacity, the cooling rate can be well approximated by  $\mathcal{L_{\rm LTE}}(\tau,\tau_{\rm d}) \approx \mathcal{L_{\rm LTE}}(\tau_{\rm d}~=~0) \times E_{2}(0.95\;\tau_{\rm d})$.  This allows us to investigate the maximum cooling effect $\water$ line cooling will have in chondrule-forming nebular shocks. 

Because vibrational line photons from $\water$ so effectively cool the gas, the high oxygen fugacities and $\water$ densities 
inferred for the chondrule-forming environment (Fedkin \& Grossman 2007) would seem incompatible with 
cooling rates $<$ 100 K/hr, but we find that dust significantly reduces the photon escape probabilities 
and the cooling rates.  
Dust grains absorb the line photons that would escape and cool the gas.  
Without dust, $\water$ line cooling reduces the temperature of the gas to 1400 - 1800$\;$K in $\leq$ 
0.1 hr and cooling continues at this rate ($>$10$^4$ K/hr).  
With dust, however, the grains absorb line photons and inhibit cooling, 
leading to a cessation of cooling  after $\sim$ 0.1 hr ($\sim$ 8 min).  Line cooling is a very effective and important cooling agent in the first few minutes and must be included in any comprehensive shock model. 
After the first few minutes, however, the thermally well-coupled gas and chondrules cool together at a slow rate (only as fast as they travel $\sim$ 1 optical depth through 
the dust), as in DC02 and CH02.  INSN failed to note this effect;  although they treat the gas and dust as well coupled (T$_{\rm gas}$ = T$_{\rm dust}$), their system was effectively optically thin to line emission, allowing all line photons to escape the region and cool the gas.    

As we are interested at this time only in the maximum cooling effects of line photons (the ``worst case scenario"), our calculations of cooling rates made no assumptions about radiative transfer; the chondrule temperatures followed the gas temperature.  
Thermal histories of chondrules including radiative transfer, line cooling, and other relevant effects will be reported on in 
future work (Morris et al.\ 2009b, in preparation).  
Since we have assumed that line photons either escape completely or are 
absorbed immediately, the true situation will clearly be bracketed between the two extremes shown in 
Figure~\ref{fig:chondrule}. 
In these calculations, we have assumed that the dust is thermodynamically coupled to the gas.  
All temperatures (gas, dust, and chondrules), in actuality, will be determined by dust/chondrule opacity and radiative 
transfer.  
DC02 showed that the cooling rates of chondrules (without the inclusion of line cooling) were
given by the formula that appears above as Equation 1. 
In the case considered in Figure~\ref{fig:chondrule}, the chondrule concentration may be considered to be low 
($C \ll 1$), and the dust opacity is characterized by 
$\delta = 0.1 \times 28.76 / 1.14 = 2.5$.
This yields a total cooling rate of gas and dust through the crystallization range of about
$125 \, {\rm K} \, {\rm hr}^{-1}$ after line cooling shuts off. 
Higher chondrule concentrations ($C > 10^2$) appear to have been typical during 
chondrule formation (Cuzzi \& Alexander 2006; Alexander et al.\ 2008).
These would increase the cooling rate to several $\times 10^{2} \, {\rm K} \, {\rm hr}^{-1}$.
Line cooling is significant during the first few minutes past the shock, and more 
detailed calculations will be necessary to constrain the initial drop in temperature 
from the peak to below the liquidus.
However, our results have shown that dust effectively shuts down line cooling within 
minutes and it then 
takes hours for chondrules to crystallize in a shock, consistent with DC02 and CH02.  

As mentioned previously, we did not calculate line cooling due to CO, as it is a factor of \app 20 less 
effective than $\water$.  
CO also cools by emitting line photons, but dust will absorb these line photons as readily as those 
emitted by $\water$.  
Figure~\ref{fig:chondrule} shows a drop in temperature of \app 600~K.  
If CO were included, the temperature would probably drop an extra 5\% (\app 630~K), which is within the 
uncertainties of the $\water$ abundances we have assumed.   

This study has shown that the gas and chondrules will cool much too rapidly 
to match the experimental 
constraints on chondrule cooling rates if dust is not present.  
The question of the dust abundance and its opacity therefore becomes paramount.  But, significantly, we have found that enhancing only the amount of $\water$ causes higher cooling rates, whereas enhancing 
the amount of $\water$ and dust together does not.  What affects the cooling rate the most, therefore, 
is not the amount of water, but the dust-to-water ratio.  Potential complications, though, include the possibility that dust will evaporate immediately upon passage 
of the shock, and the possibility that dust vapor may recondense in the post-shock region
(e.g., Scott \& Krot 2005).

In a related vein, we note an application of this work only tangentially related to chondrule formation.
Ciesla et al.\ (2003) suggested that phyllosilicates found in the fine-grained accretionary rims of 
chondrules in CM meteorites may have resulted from rapid gas-phase reactions of water vapor with silicates.
Previous studies indicated that phyllosilicate production would be kinetically inhibited in the solar 
nebula (Prinn \& Fegley 1987); however, Ciesla et al. (2003) showed that nebular shocks could result in 
a local increase in the water vapor pressure, thereby increasing the rate of phyllosilicate formation, as 
well as increasing the temperature at which phyllosilicates become stable.  
Chondrule-forming shocks in icy regions of the solar nebula could therefore account for both the formation 
of chondrules and their fine-grained phyllosilicate rims.
A possible objection to this hypothesis is that the high abundance of water vapor in the shocked region
would cool the gas too rapidly to allow the production of much phyllosilicates.
As our work here shows, however, rapid cooling by line photons will not be associated with shocked,
water-rich nebula gas, as long as the dust-to-water ratio is close to the canonical value for the solar
nebula. 
An investigation of this hypothesis is also planned for future work.  

\acknowledgements This work was supported by NASA Origins of Solar Systems grant NNG06GI65G.  We would like to thank Frank Timmes for assistance on certain computing aspects of the problem.

\appendix

\section{Line Cooling and Optical Depth}

The cooling per line is given by the expression

\begin{equation}
\frac{\Lambda{ul}}{n_{H_{2}O}} = \frac{n_u}{n_{H_{2}O}} \; A_{ul} \; h\nu_{ul}, 
\label{eq23}
\end{equation}

\bigskip

\noindent where (in LTE) 

\begin{equation}
\frac{n{u}}{n_{H_{2}O}} = \frac{n_l}{n_{H_{2}O}} \; \frac{g_u}{g_l} \; e^{-{h\nu}/{kT}}, 
\label{eq24}
\end{equation}

\bigskip

\noindent and the Einstein coefficient, $A_{ul}$ is given by

\begin{equation}
A_{ul} = \frac{2h\nu^3}{c^2} \; \frac{B_{lu}g_l}{g_u}. 
\label{eq25}
\end{equation}

\bigskip

\noindent Substituting Eq. \ref{eq24} and Eq. \ref{eq25} into Eq. \ref{eq23} gives

\begin{eqnarray}
\frac{\Lambda{ul}}{n_{H_{2}O}} &=& \left( \frac{n_l}{n_{H_{2}O}} \; e^{-{h\nu}/{kT}} \right)  \; \left( \frac{2h\nu^3}{c^2} \; B_{lu} \right) \; h\nu \nonumber \\
&=& \frac{n_l}{n_{H_{2}O}} \; \frac{h\nu}{4\pi} \; B_{lu} \left( \frac{8\pi h\nu^3}{c^2} \; e^{-{h\nu}/{kT}} \right) \nonumber \\
&=& \frac{n_l}{n_{H_{2}O}} \; \frac{h\nu}{4\pi} \; B_{lu} \; \left( 1 - e^{-{h\nu}/{kT}} \right) \left( \frac{8\pi h\nu^3}{c^2} \; \frac{e^{-{h\nu}/{kT}}}{1 - e^{-{h\nu}/{kT}}} \right) \nonumber\\
&=& \frac{n_l}{n_{H_{2}O}} \; \frac{h\nu_{lu}}{4\pi} \; B_{lu} \; \left( 1 - e^{-{h\nu}/{kT}} \right) \left(4 \pi B_{\nu}(T) \right)
\label{eq26}
\end{eqnarray}

\bigskip

What we are given in the SCAN-\water\ line list is linestrength.  The 1.2 million lines of spectral data included in the line list cover a range in wavelength from $\sim$~6700~$\mbox{\AA}$ to 25~$\mu$m.  The data consist of the following: the wavenumber at line center, $\nu_{0}$, in cm$^{-1}$, the temperature-independent line strength, $S_0$, in km/mol, and the excitation energy, $E_{low}$, in cm$^{-1}$.  From the temperature-independent line strength, $S_0$, given in the list, the temperature-dependent linestrength, $S(T)$, is calculated as follows:\\

\begin{equation}
S(T)=\frac{S_0\:\exp\left(\:-E_{low} \frac{h c}{k T}\right) \left(1-\exp\:\left(-\nu_0 \frac{h c}{k T}\right)\right)}{Q_{vib}(T)Q_{rot}(T)}.
\label{eq7} 
\end{equation}\\

The vibrational partition function, $Q_{vib}$, is given for temperatures ranging from 200 K to 8000 K, in increments of 200 K, in the documentation for the SCAN-H$_2$O line list.  We have interpolated values for $Q_{vib}$ using a quadratic spline.  The rotational partition function, $Q_{rot}$, is calculated with a subroutine provided in the documentation for the SCAN-H$_2$O line list.  $S(T)$, as shown by Eq. \ref{eq7}, is in units of km/mol, and we wish to have $S(T)$ in cm$^2$~$\cdot$~Hz, so we must convert wavenumber to frequency which gives

\begin{equation}
S(T,\nu)=10^{5}\:\cdot\:S_{0}\:\frac{c}{N_A},
\label{eq9.1}
\end{equation}\\

\noindent where $N_{A}$ is Avogadro's number, and we have converted from km to cm. This results in a conversion factor where

\begin{equation}
S(T,\nu)=\mbox{4.98 x 10}^{-9}\:S_0=S(T).
\label{eq10}
\end{equation}\

The left side of Eq. \ref{eq26} is just $S(T)$, as given by Eq. \ref{eq7}, so in terms of linestrength

\begin{eqnarray}
\frac{\Lambda{ul}}{n_{H_{2}O}} &=& S(T) \left( \frac{8\pi h\nu^3}{c^2} \; \frac{1}{e^{-{h\nu}/{kT}}-1} \right) \nonumber \\
&=& S(T) \left( 8\pi \; \frac{\nu^2}{c^2} \; kT \; \frac{h\nu/kT}{e^{-{h\nu}/{kT}}-1} \right) \nonumber \\
&=& S(T) \left( \frac{8\pi}{\lambda^2} \; kT \; \frac{h\nu/kT}{e^{h\nu/kT} - 1} \right).
\label{eq27}
\end{eqnarray}

\bigskip

According to Plume et al. (2004), the optical depth to line center is given by

\begin{equation}
\tau_{0} = N_l\;\frac{g_u}{g_l}\;\frac{A_{ul}}{\sqrt{\pi}}\;\frac{\lambda^3}{8\pi}\;\ \frac{\left( 1-e^{-{h\nu}/ kT}\right)}{\Delta v},
\label{eq17}
\end{equation}

\noindent where $\Delta v = \left(2kT/m \right)^{1/2}$ and the Einstein coefficients, $A_{ul}$ and $B_{lu}$ are given as follows: 

\begin{equation}
A_{ul} = \frac{2h\nu^3}{c^2}\;B_{ul}\; = \;\frac{2h\nu^3}{c^2}\;\frac{g_l}{g_u}\;B_{lu},
\label{eq18}
\end{equation}

\noindent which gives the gas optical depth as

\begin{equation}
\tau_0 = N_l \; \frac{1}{\sqrt{\pi}} \; \frac{1}{8\pi} \; \frac{c^3}{\nu^3}\;\frac{2h\nu^3}{c^2} \; \;\frac{B_{lu}}{\Delta v}\left( 1-e^{-{h\nu}/{kT}}\right).
\label{eq19}
\end{equation}\

\noindent In local thermodynamic equilibrium (LTE),

\begin{equation}
N_l=N_{H_{2}O} \; \frac{g_l \; \exp{\left({-E_{low}}/{kT}\right)}}{Q_{vib}(T)Q_{rot}(T)}, 
\label{eq19.1}
\end{equation}

\noindent which gives

\begin{equation}
\tau_0=N_{H_{2}O} \; \frac{g_l \; \exp{\left({-E_{low}}/{kT}\right)}}{Q_{vib}(T)Q_{rot}(T)} \; \frac{1}{4\pi} \; \frac{hc}{\sqrt{\pi}} \; \frac{B_{lu}\left(1-e^{-{h\nu}/{kT}}\right)}{\Delta v}.
\label{eq19.2}
\end{equation}

\noindent  Integrating over all frequencies, and normalizing to $\Delta \nu_0 = \Delta v/c \; \nu_0$ gives the temperature-dependent line strength

\begin{equation}
S_{ul} = S(T) = \frac{n_l}{n_{H_{2}O}} \; \frac{h\nu}{4\pi} \; B_{lu} \left(1-e^{-{h\nu}/{kT}}\right),
\label{eq20}
\end{equation}\

\noindent where the units are cm$^2$ Hz molecule$^{-1}$, resulting in the gas optical depth to line center

\begin{equation}
\tau_{0} = N_{H_{2}O} \; \frac{S_{ul}}{\sqrt{\pi}\left(\Delta v/c \right) \nu_{0}}.
\label{eq21}
\end{equation}

\bigskip

\noindent  We want the frequency-integrated optical depth, however, which is given by 

\begin{equation}
\tau_{0} = N_{H_{2}O} \; \frac{S_{ul}}{\left(\Delta v/c \right)\nu_{0}}.
\label{eq22}
\end{equation}

\bigskip

\end{document}